\documentclass[a4paper,12pt]{article}

\usepackage[utf8]{inputenc}
\usepackage{comment}
\usepackage[margin=2.5cm]{geometry}
\usepackage{graphicx}
\usepackage{caption}
\usepackage{subfig}
\usepackage{hyperref}
\usepackage{amsmath}
\usepackage{xcolor}
\usepackage{authblk}

\usepackage{multirow}

\title{Quasinormal mode spectrum of rotating black holes in Einstein-Gauss-Bonnet-dilaton theory}

\author[1]{Jose Luis Bl\'azquez-Salcedo\thanks{\href{mailto:jlblaz01@ucm.es}{jlblaz01@ucm.es}}} 
\author[1,2]{Fech Scen Khoo\thanks{\href{mailto:fkhoo@ucm.es}{fkhoo@ucm.es}}}
\author[2]{Burkhard Kleihaus\thanks{\href{mailto: b.kleihaus@uni-oldenburg.de}{b.kleihaus@uni-oldenburg.de}}} 
\author[2]{Jutta Kunz\thanks{\href{mailto:jutta.kunz@uni-oldenburg.de}{jutta.kunz@uni-oldenburg.de}}}  
\affil[1]{Departamento de F\'isica Te\'orica and IPARCOS, Facultad de Ciencias F\'isicas, Universidad Complutense de Madrid, 28040 Madrid, Spain}
\affil[2]{Institut f\"ur  Physik, Universit\"at Oldenburg, Postfach 2503,
D-26111 Oldenburg, Germany}

\date{\today}

\begin{document}

\maketitle

\begin{abstract}
Quasinormal modes are excited during the ringdown phase of black holes after merger.
Determination of quasinormal modes of rapidly rotating black holes in alternative theories of gravity has remained a challenge for a long time.
Here we discuss in detail our recently developed method to extract the quasinormal modes for rapidly rotating black holes in Einstein-Gauss-Bonnet-dilaton theory.
We first obtain numerically the exact rapidly rotating background solutions, which also clarify their domain of existence.
Then we solve the equations for the linear perturbations of the metric and the dilaton field by employing an appropriate set of boundary conditions and a spectral decomposition of the perturbation functions.
The resulting spectrum agrees well with the known limits obtained for slow rotation and weak coupling, while it exhibits larger deviations for stronger coupling.
\end{abstract}

\section{Introduction}

The quest for a more fundamental theory of gravity beyond General Relativity has sparked numerous directions of investigation and led to a rich variety of alternative theories of gravity (see e.g.\cite{Faraoni:2010pgm,Berti:2015itd,CANTATA:2021ktz}).
While any viable theory needs to fulfill solar system constraints, the strong gravity sector is still far less constrained \cite{Will:2018bme}.
Thus current and future observations of compact objects in the strong gravity regime are of high significance to place bounds on gravitational theories or even rule them out, by comparing with theoretical predictions obtained in these theories.

Here we focus on gravitational waves from the merger of compact objects and, in particular, black holes, as observed by the LIGO/Virgo/KAGRA collaboration \cite{LIGOScientific:2016aoc,KAGRA:2013rdx,Cahillane:2022pqm} and future gravitational wave detectors (see e.g. \cite{Punturo:2010zz,Dwyer:2015,Colpi:2024xhw}).
In these violent processes the gravitational wave emission features three phases, inspiral, merger and ringdown, where each phase has its own signatures.
In the final phase, the ringdown phase, the newly formed compact object emits a set of gravitational modes, that are characteristic of the object and the underlying theory, namely its associated quasinormal modes \cite{Kokkotas:1999bd,Berti:2009kk,Konoplya:2011qq}.

The determination of quasinormal modes of rotating compact objects is far more challenging than in the case of static ones.
Making use of the Newman-Penrose formalism, the Teukolsky equation was derived long ago in order to extract the quasinormal modes of the Kerr black holes of General Relativity (GR) \cite{Teukolsky:1973ha}.
The calculation of the quasinormal modes of the Kerr-Newman black holes of GR, however, was only achieved much more recently and required numerical methods to solve a system of two coupled partial differential equations (PDEs) \cite{Dias:2015wqa,Dias:2021yju,Dias:2022oqm}.
Thus in alternative theories of gravity numerical methods should also be typically necessary to extract the quasinormal modes of rapidly rotating black holes.
Nonetheless there has been progress made for higher derivative gravity theories \cite{Cano:2023tmv,Cano:2023jbk,Cano:2024ezp},
and scalar and vector modes 
were calculated under an effective field theory framework
\cite{Miguel:2023rzp}
and in a braneworld scenario \cite{Bohra:2023vls}.

A well-motivated interesting class of alternative theories of gravity are the Einstein-Gauss-Bonnet-scalar theories (EGBs).
These theories may be seen from an effective field theory point of view, since they feature quadratic gravity terms, that arise in a derivative expansion of the metric \cite{Burgess:2003jk}.
In EGBs theories the coefficients of the quadratic terms are chosen to precisely yield the Gauss-Bonnet (GB) invariant, that is present in the higher-dimensional Lovelock theories, but does not contribute to the equations of motion in four dimensions.
Here it is the coupling to an additional scalar degree of freedom, that leads to non-trivial modifications of the Einstein equations and makes EGBs theories a subset of Horndeski theories \cite{Horndeski:1974wa}.

On the other hand, string theory also predicted EGBs theories to arise in the low-energy limit \cite{Gross:1986mw,Metsaev:1987zx}.
In this case the scalar field is the dilaton, and it is coupled to the GB term with an exponential coupling function, $f(\varphi)=\exp{(-\varphi)}$.
For a finite GB coupling constant $\alpha$ the theory no longer admits the black holes of GR, instead hairy black holes arise \cite{Kanti:1995vq,Torii:1996yi,Guo:2008hf,Pani:2009wy,Kleihaus:2011tg,Pani:2011gy,Ayzenberg:2013wua,Ayzenberg:2014aka,Kleihaus:2014lba,Maselli:2015tta,Kleihaus:2015aje,Blazquez-Salcedo:2016enn,Cunha:2016wzk,Zhang:2017unx, Blazquez-Salcedo:2017txk,Pierini:2021jxd,Pierini:2022eim}.
The domain of existence of Einstein-Gauss-Bonnet-dilaton (EGBd) black holes is limited and yields a theoretical bound on the GB coupling constant in terms of the black hole mass \cite{Kanti:1995vq}.

The quasinormal modes of the static EGBd black holes were obtained earlier \cite{Blazquez-Salcedo:2016enn,Blazquez-Salcedo:2017txk}, showing already that isospectrality is no longer present due to the presence of the scalar field.
Nevertheless, for small coupling $\alpha$ the deviations from Schwarzschild remained small, and increased only substantially towards the boundary of the domain of existence.
Subsequent studies performed perturbative calculations of the quasinormal modes in both the coupling constant and the angular momentum \cite{Pierini:2021jxd,Pierini:2022eim}.

Here we present our detailed study of the spectrum of quasinormal modes of rapidly rotating EGBd black holes, where both the coupling and the rotation are treated fully non-perturbatively \cite{Blazquez-Salcedo:2024oek}.
We thus start by recalling the the exact background solutions in section 2. 
We discuss the metric and scalar perturbations together with the applied numerical spectral method in section 3.
We present the results for the spectrum in section 4 and conclude in section 5.

\section{Rotating EGBd black holes}
\label{setup}

We now present the action and field equations of EGBd theory and briefly discuss the rapidly rotating EGBd black holes \cite{Blazquez-Salcedo:2024oek} and their properties. 
In particular, we illustrate the domain of existence of EGBd black holes \cite{Kleihaus:2011tg,Kleihaus:2014lba,Kleihaus:2015aje}.

\subsection{Theory and field equations}

The Einstein-Gauss-Bonnet-dilaton action in four dimensions is given by
\begin{equation}
    S(g,\varphi) = \frac{1}{16\pi} \int d^4x \sqrt{-g}
    \left(R - \frac{1}{2}\partial_{\mu}\varphi \, \partial^{\mu}\varphi 
    +  \alpha e^{-\gamma\varphi} R^2_{\text{GB}}
    \right)\, ,
\end{equation}
where $R$ is the curvature scalar,
$\varphi$ is the dilaton field,
$\alpha$ is the GB coupling constant,
$\gamma$ is the dilaton coupling constant, and $R^2_{\text{GB}}$
is the Gauss-Bonnet term,
\begin{equation}
    R^2_{\text{GB}} = R_{\mu\nu\rho\sigma}R^{\mu\nu\rho\sigma}
- 4 R_{\mu\nu}R^{\mu\nu} + R^2 \,.
\end{equation}

Variation of the action leads to the coupled set of field equations for the metric and the scalar field.
The metric field equations may be written in the following form
\begin{equation}
    G_{\mu\nu} 
    - \frac{1}{2}T_{\mu\nu}^{(\varphi)}
    + \alpha e^{-\gamma\varphi}
    \left[H_{\mu\nu}+ 4(\gamma^2 \nabla^{\rho} \varphi \nabla^{\sigma}\varphi - \gamma \nabla^{\rho}\nabla^{\sigma} \varphi) P_{\mu\rho\nu\sigma} \right] = 0 \, ,
\end{equation}
where
\begin{eqnarray}
    G_{\mu\nu} &=& R_{\mu\nu} - \frac{1}{2}g_{\mu\nu} R \, ,\nonumber \\
    T_{\mu\nu}^{(\varphi)} &=&
    \nabla_{\mu}\varphi \nabla_{\nu} \varphi
    - \frac{1}{2} g_{\mu\nu}(\nabla\varphi)^2 \,,
    \nonumber \\
    H_{\mu\nu} &=& 2 (RR_{\mu\nu} - 2R_{\mu\rho} R^{\rho}{}_{\nu} - 2R^{\rho\sigma}R_{\mu\rho\nu\sigma} +
    R_{\mu}{}^{\rho\sigma\lambda} R_{\nu\rho\sigma\lambda} ) - \frac{1}{2}g_{\mu\nu} R^2_{\text{GB}} \, ,
     \nonumber \\
     P_{\mu\nu\rho\sigma} &=& R_{\mu\nu\rho\sigma}
     + 2 g_{\mu[\sigma} R_{\rho]\nu} 
     + 2  g_{\nu[\rho} R_{\sigma]\mu}
     + R g_{\mu[\rho} g_{\sigma]\nu} \,,
\end{eqnarray}
and the dilaton equation reads
\begin{eqnarray}
\label{dil-eq}
\nabla^2 {\varphi} -\alpha \gamma e^{-{\gamma \varphi}}  R^2_{\rm GB}
 =0.
\end{eqnarray} 
In the following we choose for the dilaton coupling constant the value $\gamma=1$ \cite{Kleihaus:2011tg,Kleihaus:2014lba,Kleihaus:2015aje}.

\subsection{Stationary rapidly rotating background}

The metric functions of the stationary axially symmetric black hole background spacetimes depend only on the radial coordinate $r$ and the polar angle $\theta$.
As previously \cite{Kleihaus:2011tg,Kleihaus:2014lba,Kleihaus:2015aje} we work here with a quasi-isotropic radial coordinate. 
The metric can then be expressed as follows
\begin{eqnarray}
ds^2 = 
g_{tt}(r,\theta) dt^2  
+ g_{rr}(r,\theta) dr^2  
+ g_{\theta\theta}(r,\theta) d\theta^2  + g_{\phi\phi}(r,\theta) d\phi^2  + g_{t\phi}(r,\theta) dt d\phi
\, ,
\label{metric_1}
\end{eqnarray}
with metric coefficients
\begin{eqnarray}
   -g_{tt}(r,\theta) = f(r,\theta) - \frac{l(r,\theta)}{f(r,\theta)}w^2(r,\theta)\sin^2{\theta} \, , \\
   g_{rr}(r,\theta) = \frac{m(r,\theta)}{f(r,\theta)} \, , \\
   g_{\theta\theta}(r,\theta) = \frac{m(r,\theta)}{f(r,\theta)}r^2 \, , \\
   g_{\phi\phi}(r,\theta) = \frac{l(r,\theta)}{f(r,\theta)}r^2\sin^2{\theta} \, , \\
   g_{t\phi}(r,\theta) = -\frac{l(r,\theta)}{f(r,\theta)}w(r,\theta) r \sin^2{\theta}\, . 
\end{eqnarray}
Let us first address the case of a vanishing scalar field.
When the Kerr metric is parametrized analogously in quasi-isotropic coordinates instead of the usual Boyer-Lindquist coordinates, the metric functions $f(r,\theta)$, $l(r,\theta)$, $m(r,\theta)$, and $w(r,\theta)$ read \cite{Kleihaus:2015aje}
\begin{eqnarray}
f(r,\theta) = \left( 1 - \frac{r_H^2}{r^2} \right)^2 \frac{F_1(r,\theta)}{F_2(r,\theta)}
, \\
l(r,\theta) = \left( 1 - \frac{r_H^2}{r^2} \right)^2 
, \\
m(r,\theta) = \left( 1 - \frac{r_H^2}{r^2} \right)^2 \frac{F_1^2(r,\theta)}{F_2(r,\theta)}
, \\
w(r,\theta) = \frac{2J}{r^2F_2(r,\theta)}\left( 1+ \frac{M}{r} + \frac{r_H^2}{r^2} \right)
, \\
  F_1(r,\theta) = \frac{2M^2}{r^2} + \left( 1 - \frac{r_H^2}{r^2} \right)^2
   + \frac{2M}{r}\left( 1 + \frac{r_H^2}{r^2} \right)
  - \frac{J^2}{M^2r^2}\sin^2{\theta}
  , \\
  F_2(r,\theta) = \left(\frac{2M^2}{r^2} + \left( 1 - \frac{r_H^2}{r^2} \right)^2
   + \frac{2M}{r}\left( 1 + \frac{r_H^2}{r^2} \right)\right)^2
  - \left( 1 - \frac{r_H^2}{r^2} \right)^2\frac{J^2}{M^2r^2}\sin^2{\theta}
  , 
\end{eqnarray}
and the scalar field vanishes, of course.
{$r_H$ is the location of the horizon, $M$ is the total mass, and $J$ is the total angular momentum.}
Furthermore, the following relations hold for Kerr ($\alpha=0$),
\begin{eqnarray}
   \Omega_H =  w(r_H,\theta)/r_H = \frac{\sqrt{M^2-4r_H^2}}{2M(M+2r_H)} \, , \\
    J = M\sqrt{M^2-4r_H^2} \, .
\end{eqnarray}

Next we consider the general case of a non-vanishing scalar field. First we discuss the 
behavior of the metric functions and the scalar field in the asymptotic region, at the horizon, and the boundary conditions. Then we
turn to the numerical construction of the global EGBd black holes.
Since we are interested in asymptotically flat black holes, the spacetime has to approach Minkowski space for $r\to \infty$.
The asymptotic expansion then yields
\begin{eqnarray}
   -g_{tt}(r,\theta) \approx 1 - \frac{2M}{r} + ... \, , \\
   g_{t\phi}(r,\theta) \approx -\frac{2J}{r}\sin^2{\theta}  + ... \, ,\\
   g_{rr}(r,\theta) \approx 1 + \frac{\hat{f_1}}{r} + ... \, , \\
   g_{\theta\theta}(r,\theta) \approx r^2 + \hat{f_1}r + ... \, , \\
   g_{\phi\phi}(r,\theta) \approx r^2\sin^2{\theta} + ... \, ,  \\
   \varphi(r,\theta) \approx \frac{Q_d}{r} + ...  \, ,
\end{eqnarray}
where $Q_d$ is scalar charge and $\hat{f_1}$ is a numerical parameter.

The horizon resides at a constant value of the radial coordinate $r_H$, that is determined by the condition $f(r_H)=0$ \cite{Kleihaus:2000kg,Kleihaus:2002ee,Kleihaus:2003sh}.
Requiring the horizon to be regular, the asymptotic expansion for $r\to r_H$ of the metric functions and the scalar field becomes
\begin{eqnarray}
   g_{tt}(r,\theta) = \Omega_H^2 k_0(\theta) + k_1(\theta)(r-r_H)^2 +  O((r-r_H)^3) \, , \\
   g_{t\phi}(r,\theta) = -\Omega_H k_0(\theta) + k_2(\theta)(r-r_H)^2 +  O((r-r_H)^3)  \, , \\
   g_{rr}(r,\theta) = k_3(\theta) + O((r-r_H))  \, , \\
   g_{\theta\theta}(r,\theta) = r_H^2 k_3(\theta) +  O((r-r_H)^2)  \, , \\
   g_{\phi\phi}(r,\theta) = k_0(\theta)  + k_4(\theta)(r-r_H)^2 +  O((r-r_H)^3)   \, , \\
   \varphi(r,\theta) = k_5(\theta) + + O((r-r_H)^2)  \, , 
\end{eqnarray}
with the horizon angular velocity $w(r_H,\theta) = \Omega_H r_H$, and the numerical functions $k_i(\theta)$, $i=0,...,5$, that satisfy some complicated relations.

Furthermore, requiring regularity on the symmetry axis yields the condition, that the derivatives of the functions with respect to the polar angle $\theta$ need to vanish on the axis,
\begin{equation}
   \partial_\theta f|_{\theta={0,\pi}} =
   \partial_\theta m|_{\theta={0,\pi}} =
   \partial_\theta l|_{\theta={0,\pi}} =
   \partial_\theta \omega|_{\theta={0,\pi}} = 
   \partial_\theta \varphi|_{\theta={0,\pi}} =
   0 \, . 
\end{equation}

\begin{figure}[t!]
\begin{center}
\subfloat[]
{\includegraphics[width=8.0cm,angle=0]{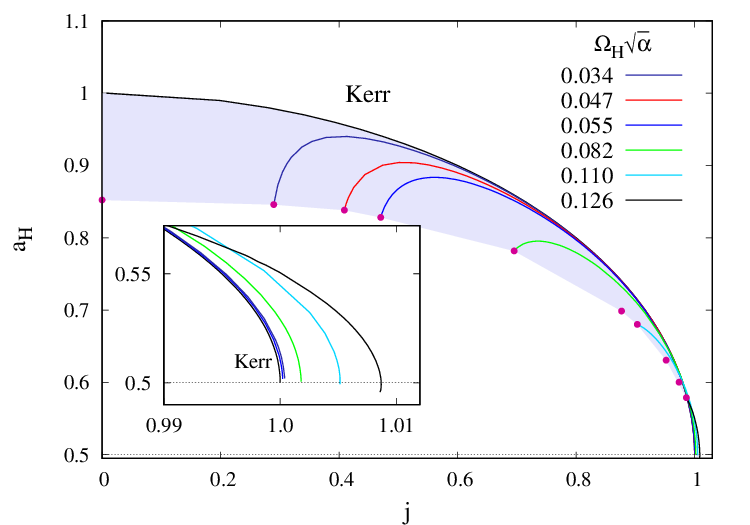}
}
\subfloat[]
{\includegraphics[width=7.9cm,angle=0]{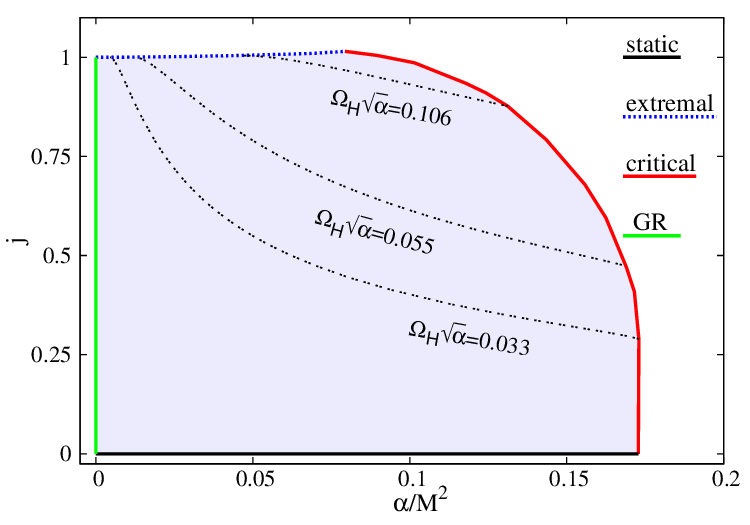} 
}
\end{center}
\caption{Rotating EGBd black holes ($\gamma=1$):
Scaled horizon area $a_{\rm H}=A_{\rm H}/16 \pi M^2$ vs scaled angular momentum $j=J/M^2$ (a)
and scaled angular momentum $j=J/M^2$ vs scaled coupling constant $\xi=\alpha/M^2$ (b).
The shaded areas indicate the respective domains of existence bounded by the Kerr, static, critical and extremal EGBd black holes.
Curves within the domains have fixed values of the scaled horizon angular velocity $\Omega_{\rm H} \sqrt{\alpha}$.
The inset (a) highlights the large $j$ region.
}
\label{fig_bg}
\end{figure} 

Now we turn to the global EGBd black hole solutions.
The resulting set of coupled PDEs were  solved numerically, using the professional package FIDISOL/CADSOL \cite{Schoenauer:1989}. The estimated numerical error for the functions is within $10^{-5}$.
For details of the numerical calculations, see \cite{Kleihaus:2011tg,Kleihaus:2014lba}.
When solving the coupled PDEs for the unknown functions subject to the boundary conditions, the domain of existence of rotating EGBd black holes can be charted \cite{Kleihaus:2011tg,Kleihaus:2014lba,Kleihaus:2015aje}.
We illustrate the domain in Fig.~\ref{fig_bg}, where we show in Fig.~\ref{fig_bg}a the scaled horizon area $a_{\rm H}=A_{\rm H}/16 \pi M^2$ versus the scaled angular momentum $j=J/M^2$.
This domain is bounded on the left by the static EGBd black holes.
Most of the lower boundary is formed by critical EGBd black holes, where a radicand vanishes, as already seen for the static black holes \cite{Kanti:1995vq}.
Most of the upper boundary is represented by the Kerr black holes.
However, the Kerr and the critical boundary cross for large $j$ and then switch roles.
Thus EGBd black holes slightly exceed the Kerr bound $j = 1$.
A last rather small part of the boundary is given by extremal EGBd black holes, as highlighted in the inset of the figure.

For our investigation of the quasinormal modes of the EGBd black holes Fig.~\ref{fig_bg}b is the most relevant.
Here we exhibit the domain of existence of the the scaled angular momentum $j=J/M^2$ versus the scaled coupling constant $\xi=\alpha/M^2$.
The boundary curves of this domain of existence are formed by the static, critical and extremal EGBd black holes as well as the Kerr black holes and do not cross but smoothly match.
The figure illustrates the maximal possible value of the scaled GB coupling for a given value of the scaled angular momentum.
We remark that beyond the Kerr bound, i.e., for $j>1$, the minimal value of the coupling is no longer zero, but greater than zero.

\section{Metric and scalar perturbations of rotating EGBd black holes}

We now discuss the metric and scalar field perturbations, present our choice of parametrization of the perturbation functions and the resulting PDEs, and then turn to the discussion of the boundary conditions and the spectral decomposition.
Thus we proceed analogously to the Kerr case, where we tested our scheme thoroughly \cite{Blazquez-Salcedo:2023hwg}, and to the case of rotating Ellis-Bronnikov wormholes \cite{Khoo:2024yeh}.

\subsection{Ansatz}
\label{sec_ansatz}

The first step is to specify the metric and the scalar field in terms of the background solutions and the perturbations.
The full metric can be written as
\begin{eqnarray}
g_{\mu\nu} &=& g^{(bg)}_{\mu\nu} + \epsilon \delta h_{\mu\nu}(t,r,\theta,\phi) \, , 
\end{eqnarray}
and the scalar field as
\begin{eqnarray}
\varphi &=& \varphi^{(bg)} + \epsilon \delta\varphi(t,r,\theta,\phi) \, , 
\end{eqnarray}
where the superscript $(bg)$ denotes the background solutions, the $\delta$ signals the perturbations, and $\epsilon$ is the perturbation parameter to keep track of the order of the perturbations.

The metric perturbations $\delta h_{\mu\nu}(t,r,\theta,\phi)$ can be further decomposed into axial and polar components, 
\begin{eqnarray}
\delta h_{\mu\nu} &=& \delta h^{(A)}_{\mu\nu} + \delta h^{(P)}_{\mu\nu}  \, ,
\end{eqnarray}
where the superscripts $(A)$ and $(P)$ denote the axial-led and polar-led perturbations.
The Ansatz for the axial and polar metric perturbations is, respectively, given by
\begin{equation}
 \delta h^{(A)}_{\mu\nu} = e^{i(M_z\phi-\omega t)} 
\begin{pmatrix}
0             & 0             & a_1(r,\theta) & a_2(r,\theta) \\
0             & 0             & a_3(r,\theta) & a_4(r,\theta) \\
a_1(r,\theta) & a_3(r,\theta) & 0             & 0 \\
a_2(r,\theta) & a_4(r,\theta) & 0             & 0
\end{pmatrix}   
\end{equation}
and
\begin{equation}
 \delta h^{(P)}_{\mu\nu} = e^{i(M_z\phi-\omega t)} 
\begin{pmatrix}
N_0(r,\theta) & H_1(r,\theta) & 0             & 0 \\
H_1(r,\theta) & L_0(r,\theta) & 0             & 0 \\
0             & 0             & T_0(r,\theta) & 0  \\
0             & 0             & 0             & S_0(r,\theta) 
\end{pmatrix}   \, .
\end{equation}
Here we have separated the dependence on time $t$ and azimuthal angle $\phi$, taking the remaining symmetries with respect to these coordinates into account.
Thus we have introduced the complex mode eigenvalue $\omega$ and the azimuthal number $M_z$ of the mode.
The eigenvalue $\omega=\omega_R + i \omega_I$ describes the frequency of a mode in terms of its real part $\omega_R$ and the damping time of a mode via the inverse of its imaginary part $\omega_I$.

For further calculations and in order to fix the gauge, it is convenient to employ the following definitions for the metric perturbation functions
\begin{eqnarray}
a_1(r,\theta) &=& - i M_z \frac{h_0(r,\theta)}{\sin{\theta}} \, , \\
a_2(r,\theta) &=& \sin{\theta} \, \partial_\theta h_0(r,\theta) \, ,  \\
a_3(r,\theta) &=& - i M_z \frac{h_1(r,\theta)}{\sin{\theta}} \, ,  \\
a_4(r,\theta) &=& \sin{\theta} \, \partial_\theta h_1 \, ,  \\
N_0(r,\theta) &=& \left( g^{(bg)}_{rr}(r,\theta) \right)^{-1}  N(r,\theta) \, ,  \\
L_0(r,\theta) &=& \left( g^{(bg)}_{rr}(r,\theta) \right)  L(r,\theta) \, ,  \\
T_0(r,\theta) &=& \left( g^{(bg)}_{\theta\theta}(r,\theta) \right) T(r,\theta)  \, , \\
S_0(r,\theta) &=& \left( g^{(bg)}_{\phi\phi}(r,\theta) \right) T(r,\theta) \, ,
\end{eqnarray}
while retaining $H_1(r,\theta)$.
The scalar field perturbation, on the other hand, is simply parameterized as
\begin{eqnarray}
\delta \varphi &=& e^{i(M_z\phi-\omega t)} \Phi(r,\theta) \, .
\end{eqnarray}
This leaves us with a set of 7 unknown perturbation functions, $H_1, T, N, L, h_0, h_1$ and $\Phi_1$.

\subsection{PDEs and parametrization in new coordinates}
\label{parametrization}

The general set of coupled equations for the metric and the scalar field may be given in the form {of}
\begin{eqnarray}
\mathcal{G}_{\mu\nu} = \mathcal{G}_{\mu\nu}^{(bg)} + \epsilon \delta\mathcal{G}_{\mu\nu} (r,\theta) e^{i(M_z\phi-\omega t)}  =0 \, , \\
\mathcal{S} = \mathcal{S}^{(bg)} + \epsilon \delta\mathcal{S} (r,\theta) e^{i(M_z\phi-\omega t)}   =0 \, .
\end{eqnarray}
Since the background solutions satisfy the equations $\mathcal{G}_{\mu\nu}^{(bg)}=0$ and $\mathcal{S}^{(bg)}=0$, the equations to be solved reduce to $\delta\mathcal{G}_{\mu\nu} (r,\theta)=0$ and $\delta\mathcal{S} (r,\theta)=0$.

We now introduce the compactified radial coordinate $x$ and the angular coordinate $y$,
\begin{eqnarray}
    x = \frac{r-r_H}{r_H+1} \, , \, \, \,   y = \cos\theta \, 
\end{eqnarray}
with domains $0 \le x \le 1$ and $-1 \le y \le 1$. 
In terms of the new radial coordinate $x=0$ corresponds to the horizon and $x=1$ to radial infinity, while the symmetry axis is located at $y=\pm 1$.

We next express the perturbation functions in terms of the new coordinates $x$ and $y$ and extract some factors that allow us to incorporate the proper boundary conditions to have purely outgoing waves at radial infinity and purely ingoing waves at the horizon. 
Our choice of isotropic coordinates for the EGBd black holes 
renders the
parametrization 
slightly
different
from the one for Kerr black holes \cite{Blazquez-Salcedo:2023hwg}.
We thus introduce the following parametrization for the metric functions
\begin{eqnarray}
 H_1 &=& \widetilde H_1(x,y)  \, \frac{1}{x(1-x)}   \, e^{i \hat{R}} \, ,
 \label{H1}
 \\
 T &=& \widetilde T(x,y)    \,  e^{i \hat{R}} \, , \\
 N &=& \widetilde N(x,y)  \, \frac{1}{1-x}    \, e^{i \hat{R}} \, , \\
 L &=& \widetilde L(x,y) \,  \frac{1}{x^2(1-x)}   \, e^{i \hat{R}} \, , \\
 h_0 &=& \widetilde h_0(x,y) \,  \frac{1}{1-x}   \, e^{i \hat{R}} \, , \\
 h_1 &=& \widetilde h_1(x,y) \,  \frac{1}{x(1-x)}   \, e^{i \hat{R}} \, 
   \label{metric_param}
\end{eqnarray}
and {for} the scalar field function
\begin{eqnarray}
\Phi = \widetilde \Phi_1(x,y)  \, (1-x)  \,  e^{i \hat{R}} \, .
\end{eqnarray}

For notational simplicity we then introduce a vector $\vec{X}$, that contains these newly defined perturbation functions, 
$\vec{X}=[\widetilde H_1, \widetilde T, \widetilde N, \widetilde L, \widetilde h_0, \widetilde h_1, \widetilde \Phi_1]$. 
In terms of this vector the resulting coupled system of 7 linear homogeneous PDEs in the   coordinates $x$ and $y$ can be compactly expressed as
\begin{eqnarray}
    \mathcal{D}_{\mathrm{I}}(x,y) \vec{X}(x,y) = 0, \, \, \, \, \quad   \mathrm{I} = 1,...,7 \, .
    \label{metric_eq_xy}
\end{eqnarray}

\subsection{Boundary conditions}

Next we specify the sets of boundary conditions more explicitly.
At radial infinity and at the horizon we here present the expansion formulae for the perturbation functions again in terms of the isotropic radial coordinate $r$ and the polar angle $\theta$ to provide more clarity.
The calculations and the compact formulations of the boundary conditions, however, employ the new set of coordinates, $x$ and $y$.

At radial infinity we require that the perturbation functions behave like a purely outgoing wave.
The perturbation functions therefore must have the well-known form
\begin{eqnarray}
           T &=& e^{i\omega R^* } \left( T^{+}(\theta) + \mathcal{O}\left(\frac{1}{r}\right) \right) \, ,\\
           H_{1} &=& r e^{i\omega R^* } \left( H^{+}_{1}(\theta) + \mathcal{O}\left(\frac{1}{r}\right) \right)\, ,\\
           N &=& r e^{i\omega R^* } \left( N^{+}(\theta) + \mathcal{O}\left(\frac{1}{r}\right)  \right) \, ,\\
           L &=& r e^{i\omega R^* } \left( L^{+}(\theta) + \mathcal{O}\left(\frac{1}{r}\right)  \right) \, ,\\
           h_{0} &=&  r e^{i\omega R^* } \left( h^{+}_{0}(\theta) + \mathcal{O}\left(\frac{1}{r}\right) \right)\, ,\\
           h_{1} &=&  r e^{i\omega R^* } \left( h^{+}_{1}(\theta)  + \mathcal{O}\left(\frac{1}{r}\right) \right)\, ,\\
           \Phi_{1} &=& \frac{1}{r} e^{i\omega R^* } \left(  \Phi_{1}^{+}(\theta) + \mathcal{O}\left(\frac{1}{r}\right)   \right)\, ,
\end{eqnarray}
where the function $R^*$ satisfies 
\begin{equation}
\frac{dR^*}{dr} =  1 + \frac{2M}{r} + \mathcal{O}\left(\frac{1}{r^2}\right) \, , 
\end{equation}
with $M$ being the total mass.

The general set of conditions can be written in operator form,
\begin{eqnarray}
    \mathcal{A}_{\mathrm{I}}(x,y) \vec{X}(x,y)|_{x=1} = 0, \, \, \, 
    \quad 
    \mathrm{I} = 1,...,7 \, ,
\label{bcg_inf}
\end{eqnarray}
where the $\mathcal{A}_{\mathrm{I}}$ are linear operators in the coordinates $x$ and $y$.

At the horizon we follow a similar procedure.
Here the perturbation functions should correspond to purely ingoing waves.
Thus we require the perturbation function to have the following expansion at the horizon
\begin{eqnarray}
           T &=& e^{-i(\omega-M_z\Omega_H) R^* } \left( T^{-}(\theta)  + 
           \mathcal{O}\left(r-r_H\right) \right)\, ,\\
           H_{1} &=& \frac{r_H}{r-r_H} e^{-i(\omega-M_z\Omega_H) R^* }  \left( H^{-}_{1}(\theta) + \mathcal{O}\left(r-r_H\right) \right)\, ,\\
           N &=&  e^{-i(\omega-M_z\Omega_H) R^* } \left( N^{-}(\theta)  + \mathcal{O}\left(r-r_H\right) \right)\, ,\\
           L &=& \frac{r_H^2}{(r-r_H)^2} e^{-i(\omega-M_z\Omega_H) R^* }  \left( L^-(\theta) + \mathcal{O}\left(r-r_H\right) \right)\, ,\\
           h_{0} &=& e^{-i(\omega-M_z\Omega_H) R^* }  \left( h^{-}_{0}(\theta) + \mathcal{O}\left(r-r_H\right) \right)\, ,\\
           h_{1} &=& \frac{r_H}{r-r_H} e^{-i(\omega-M_z\Omega_H) R^* }  \left( h^{-}_{1}(\theta) + \mathcal{O}\left(r-r_H\right) \right)\, ,\\
           \Phi_{1} &=& \frac{1}{r} e^{-i(\omega-M_z\Omega_H) R^* }  \left(  \Phi_{1}^-(\theta) + \mathcal{O}\left(r-r_H\right) \right)\, , 
\end{eqnarray}
where the function $R^*$ now satisfies 
\begin{equation}
\frac{dR^*}{dr} = \frac{g_1}{(r-r_H)} + \mathcal{O}(1) \, , 
\end{equation}
and the parameter $g_1$ is obtained from the horizon expansion of the background functions,
\begin{equation}
g_1 = \kappa_H^{-1},
\end{equation}
where $\kappa_H$ is the surface gravity.

The boundary conditions at the horizon ensuring a purely ingoing wave can also be expressed in terms of a set of linear operators $\mathcal{B}_{\mathrm{I}}$,
\begin{eqnarray}
    \mathcal{B}_{\mathrm{I}}(x,y) \vec{X}(x,y)|_{x=0} = 0, \, \, \,   \quad \mathrm{I} = 1,...,7 \, .
\label{bcg_hor}
\end{eqnarray}

We further need to require regularity on the symmetry axis, $\theta=0,\, \pi$.
On the upper half-axis $y=1$ the perturbation functions therefore need to have the regular expansion
\begin{eqnarray}
           T &=& T^{NP}(x)  + \mathcal{O}\left(y-1\right)   \, ,\\
           H_{1} &=& H_1^{NP}(x)  + \mathcal{O}\left(y-1\right)   \, ,\\
           N &=& N^{NP}(x)  + \mathcal{O}\left(y-1\right)   \, ,\\
           L &=& L^{NP}(x)  + \mathcal{O}\left(y-1\right)   \, ,\\
           h_0 &=& h_0^{NP}(x)  + \mathcal{O}\left(y-1\right)   \, ,\\
           h_1 &=&  h_1^{NP}(x)  + \mathcal{O}\left(y-1\right)   \, ,\\
           \Phi_{1} &=& \Phi_{1}^{NP}(x)  + \mathcal{O}\left(y-1\right)   \, ,
\end{eqnarray}
yielding a set of relations in an operator form
\begin{eqnarray}
    \mathbf{\alpha}_{\mathrm{I}}(x,y) \vec{X}(x,y)|_{y=1} = 0, \, \, \, 
     \quad \mathrm{I} = 1,...,7 \, .
\label{bcg_np}
\end{eqnarray}
Analogously, on the lower half-axis $y=-1$ the perturbation functions need to have the regular expansion
\begin{eqnarray}
           T &=& T^{SP}(x)  + \mathcal{O}\left(y+1\right)   \, ,\\
           H_{1} &=& H_1^{SP}(x)  + \mathcal{O}\left(y+1\right)   \, ,\\
           N &=& N^{SP}(x)  + \mathcal{O}\left(y+1\right)   \, ,\\
           L &=& L^{SP}(x)  + \mathcal{O}\left(y+1\right)   \, ,\\
           h_0 &=& h_0^{SP}(x)  + \mathcal{O}\left(y+1\right)   \, ,\\
           h_1 &=&  h_1^{SP}(x)  + \mathcal{O}\left(y+1\right)   \, ,\\
           \Phi_{1} &=& \Phi_{1}^{SP}(x)  + \mathcal{O}\left(y+1\right)   \, ,
\end{eqnarray}
yielding yet another set of relations in operator form
\begin{eqnarray}
    \mathbf{\beta}_{\mathrm{I}}(x,y) \vec{X}(x,y)|_{y=-1} = 0, \, \, \, 
     \quad \mathrm{I} = 1,...,7 \, .
\label{bcg_sp}
\end{eqnarray}

\subsection{Spectral decomposition}

We now turn to the spectral decomposition of the perturbation functions, which allows us to solve the coupled set of PDEs numerically with high accuracy, as checked in detail for the Kerr case \cite{Blazquez-Salcedo:2023hwg}.
In particular, we decompose the metric perturbations in a series of Chebyshev polynomials of the first kind $T_k(x)$ in the coordinate $x$ with $k=0,...,N_x-1$, and a series of Legendre functions of the first kind $P_l^{M_z}(y)$ in the coordinate $y$ with $l=|M_z|,...,N_y+|M_z|-1$,
\begin{eqnarray}
    \widetilde H_1(x,y) &=& \sum_{k=0}^{N_x-1}  \, \, \sum_{l=|M_z|}^{N_y+|M_z|-1} C_{1,k,l}  \,  T_k(x)  \,  P_l^{M_z}(y)   \, , \\
    \widetilde T(x,y) &=& \sum_{k=0}^{N_x-1}  \, \, \sum_{l=|M_z|}^{N_y+|M_z|-1} C_{2,k,l}  \,  T_k(x)  \, 
 P_l^{M_z}(y)   \, , \\
    \widetilde L(x,y) &=& \sum_{k=0}^{N_x-1}  \, \, \sum_{l=|M_z|}^{N_y+|M_z|-1} C_{3,k,l}  \,  T_k(x)  \,  P_l^{M_z}(y)   \, , \\
    \widetilde N(x,y) &=& \sum_{k=0}^{N_x-1}  \, \, \sum_{l=|M_z|}^{N_y+|M_z|-1} C_{4,k,l}  \,  T_k(x)  \,  P_l^{M_z}(y)   \, , \\
    \widetilde h_0(x,y) &=& \sum_{k=0}^{N_x-1}  \, \, \sum_{l=|M_z|}^{N_y+|M_z|-1} C_{5,k,l}  \,  T_k(x)  \,  P_l^{M_z}(y)   \, , \\
    \widetilde h_1(x,y) &=& \sum_{k=0}^{N_x-1}  \, \, \sum_{l=|M_z|}^{N_y+|M_z|-1} C_{6,k,l}  \,  T_k(x)  \, 
 P_l^{M_z}(y)   \, ,
    \label{metric_dec}
\end{eqnarray}
and likewise for the scalar perturbation,
\begin{eqnarray}
    \widetilde \Phi_1(x,y) &=& \sum_{k=0}^{N_x-1} \, \, \sum_{l=|M_z|}^{N_y+|M_z|-1}  C_{7,k,l}  \,  T_k(x)  \,  P_l^{M_z}(y)   \, .
\end{eqnarray}

The above double expansion introduces the set of $7 \times N_x \times N_y$ constants $C_{n,k,l}$, that are determined by solving the PDEs subject to the above specified boundary conditions. 
To that end we next need to discretize the domain of integration. 
Here we choose for the $x$-coordinate the Gauss-Lobatto points defined by
\begin{eqnarray}
x_I = \frac{1}{2} \left( 1+\cos{\left(\frac{I-1}{N_x-1}\pi\right)} \right) \, , \, \quad  I=1,...,N_x \, ,
\end{eqnarray}
and we select a uniform mesh for the $y$-coordinate
\begin{eqnarray}
   y_K = 2\frac{K-1}{N_y-1}-1 
   \, , \, \quad K=1,...,N_y \, . 
\end{eqnarray}
While there is a lot of freedom in the choice of the grid, the above grid is known to optimize the calculations in the case of Chebyshev polynomials (as seen also in our calculations of the Kerr modes \cite{Blazquez-Salcedo:2023hwg}).

In order to obtain the final set of equations, we now evaluate the corresponding boundary conditions on the boundaries of the domain of integration and evaluate the PDEs in the bulk of the domain.
This provides us with a total of $7 \times N_x \times N_y$ algebraic equations for the unknown coefficients $C_{n,k,l}$. 
These algebraic equations represent a standard quadratic eigenvalue problem, that may be expressed in matrix form
\begin{eqnarray}
    \left( \mathcal{M}_0 + \mathcal{M}_1 \omega + \mathcal{M}_2 \omega^2 \right) \Vec{C} = 0 \, ,
    \label{matrix_eq}
\end{eqnarray}
where the vector $\vec{C}$ consists of all $C_{n,k,l}$ with $n=1,...,7$, and the matrices $\mathcal{M}_0$, $\mathcal{M}_1$ and $\mathcal{M}_2$ have size $(7 \times N_x \times N_y) \times (7 \times N_x \times N_y)$.
The quasinormal modes and the eigenvectors are then obtained with the same numerical methods described previously \cite{Blazquez-Salcedo:2023hwg}, making use of Maple and Matlab with the Toolbox Advanpix \cite{Advanpix}.

\section{Spectrum of quasinormal modes for rotating EGBd black holes}

We now present our results for the spectrum of quasinormal modes of the rotating EGBd black holes.
We have focused our calculations on the modes with azimuthal number $M_z=2$ and the lower multipole numbers $l$, starting with the quadrupole $l=2$.
The label $l$ of a mode only signifies the dominant multipole contribution of a mode, since rotation mixes the multipoles.
Therefore we sometimes refer to the modes also as $l$-led modes.
Moreover we have restricted our systematic studies to the fundamental modes, leaving the overtones still unexplored.
For vanishing GB coupling $\alpha$ the modes reduce to those of the Kerr black holes.
Thus the polar-led and axial-led modes become degenerate, since the Kerr black holes feature isospectrality, and the scalar-led modes reduce to the modes of a test scalar field in the Kerr background.
This $\alpha=0$ limit together with the $j=0$ limit define our nomenclature of the modes.

\begin{figure}[t!]
\begin{center}
\includegraphics[width=6.5cm,angle=-90]{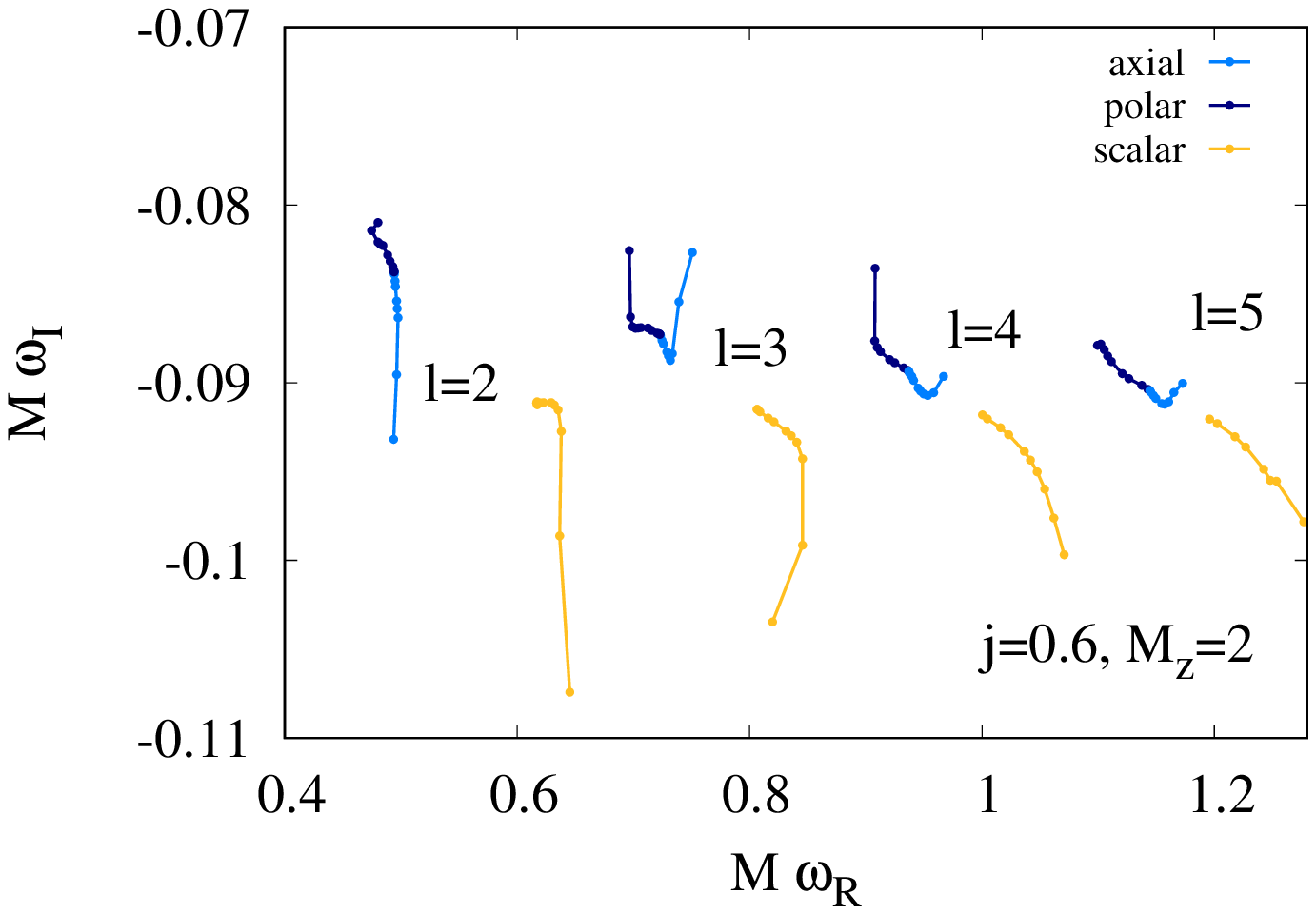}
\end{center}
\caption{Example of the quasinormal mode spectrum for EGBd black holes with scaled angular momentum $j=0.6$ and $M_z=2$, showing the fundamental modes for multipole numbers $l=2,...,5$.
The colors indicate the polar-led, axial-led and scalar-led modes.}
\label{fig_ws} 
\end{figure} 

We show in Fig.~\ref{fig_ws} an example of the quasinormal mode spectrum for rotating EGBd black holes.
The spectrum shows the scaled imaginary part $M \omega_I$ versus the scaled real part $M \omega_R$ for the fundamental quasinormal modes with azimuthal number $M_z=2$ and multipole numbers $l=2$, 3, 4, and 5, for scaled angular momentum $j=0.6$ and positive frequency.
The colors indicated the three mode types, the polar-led modes, the axial-led modes and the scalar-led modes.
The figure demonstrates the breaking of isospectrality in EGBd, as the polar-led and axial-led modes bifurcate.
Furthermore, the overall increment of the absolute value of the scaled imaginary part of the modes as $l$ grows 
indicates a mode stability of the spacetime.
A preliminary study indicates that the first overtones of the polar-led, axial-led and scalar-led modes typically possess imaginary parts with $M\omega_I<-0.2$. 

\begin{figure}[h!]
\begin{center}
\subfloat[]
{\includegraphics[width=5.5cm,angle=-90]{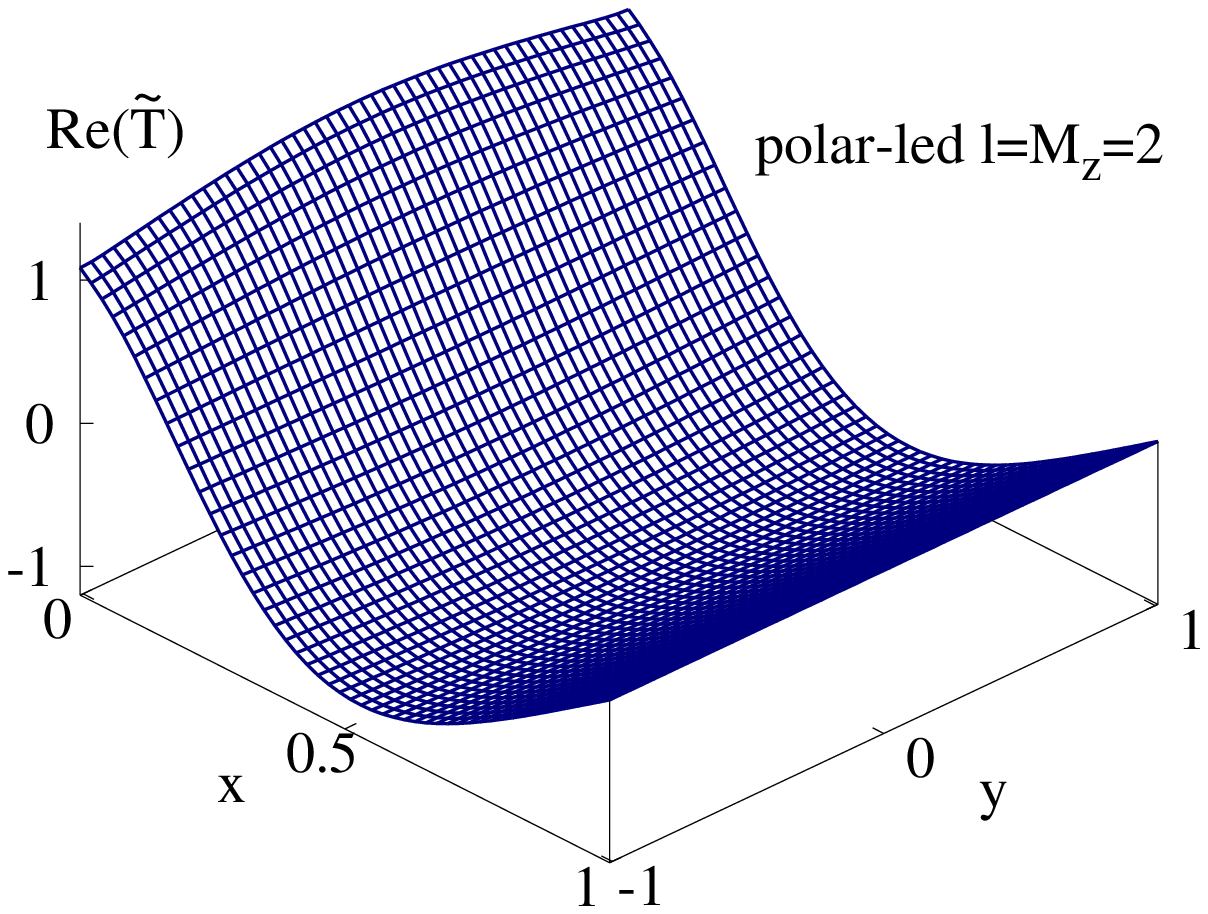}}
\subfloat[]{
\includegraphics[width=5.5cm,angle=-90]{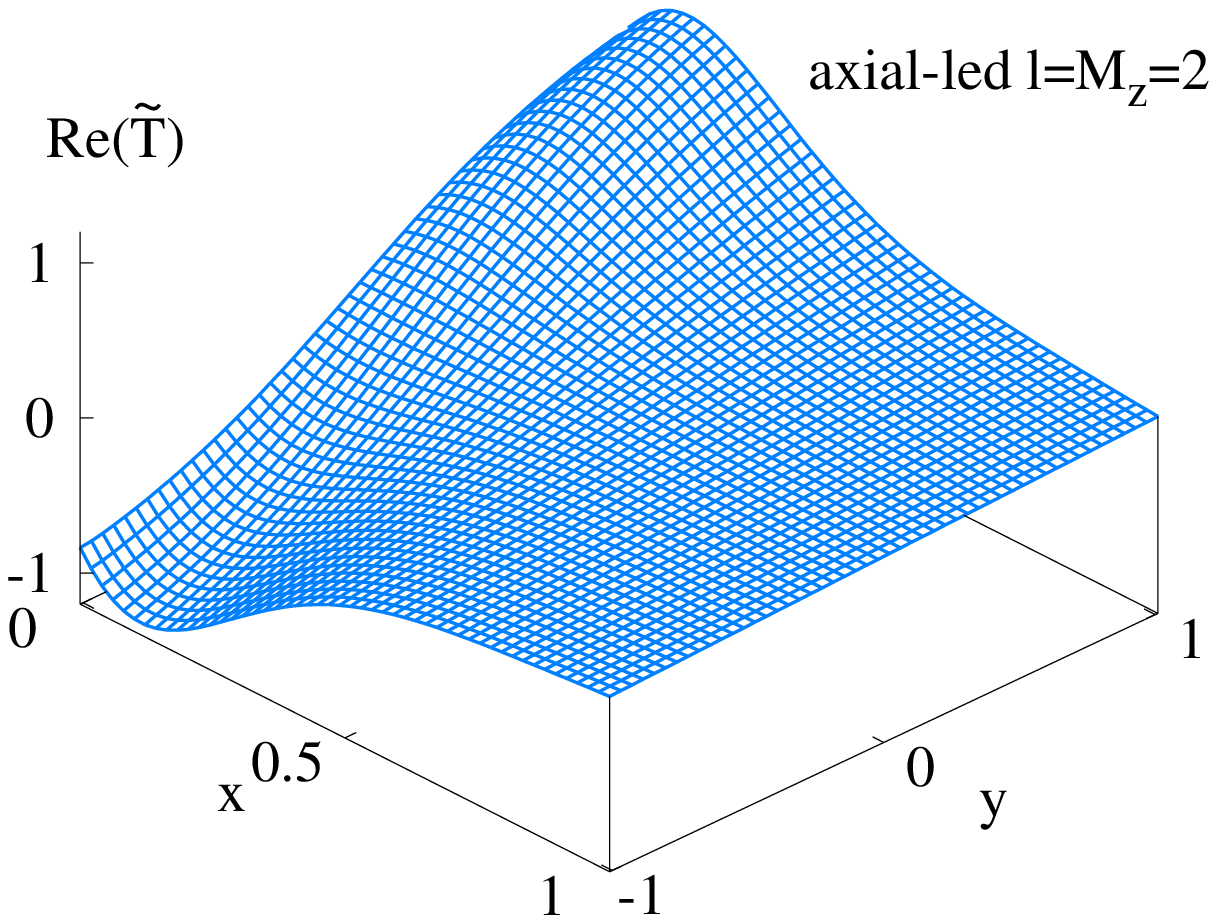}}

\subfloat[]{
\includegraphics[width=5.5cm,angle=-90]{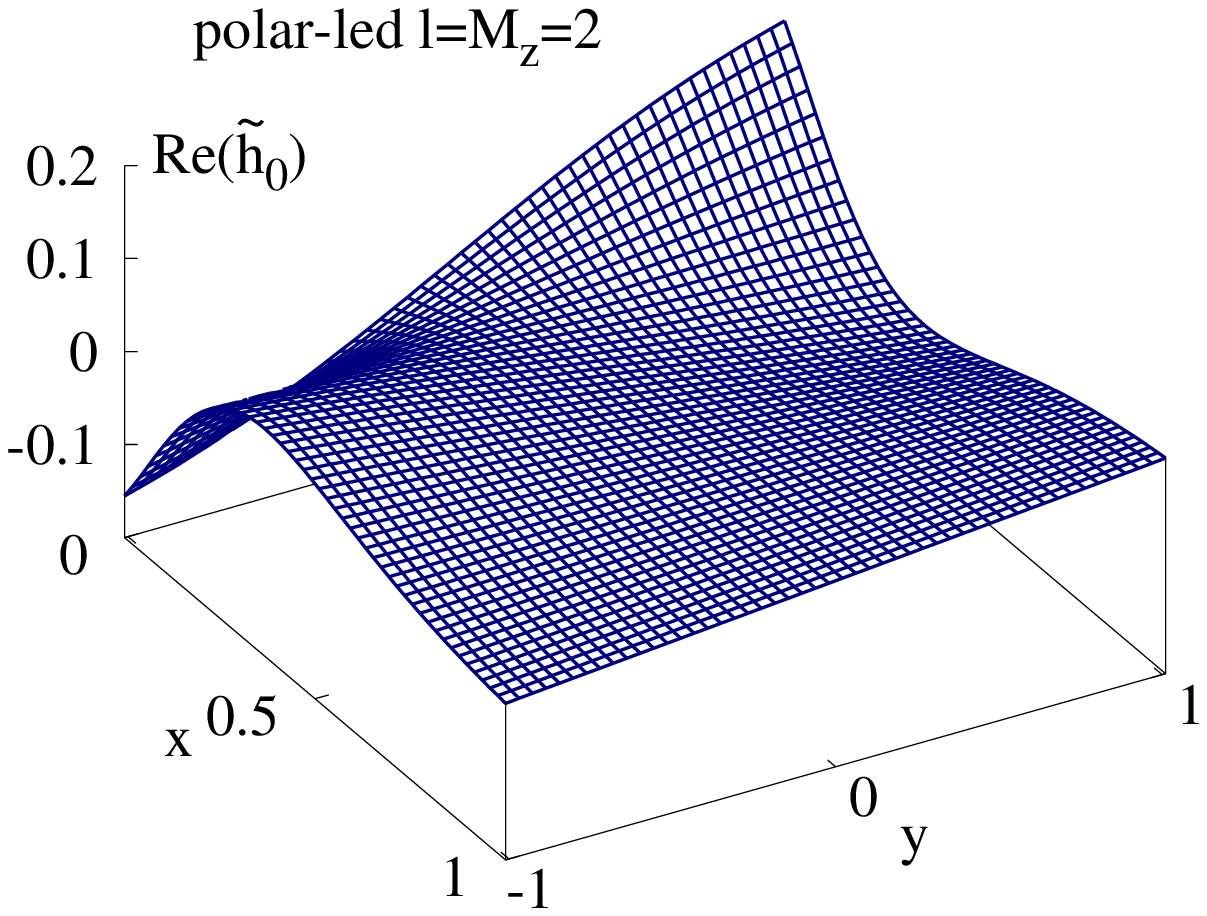}}
\subfloat[]{
\includegraphics[width=5.5cm,angle=-90]{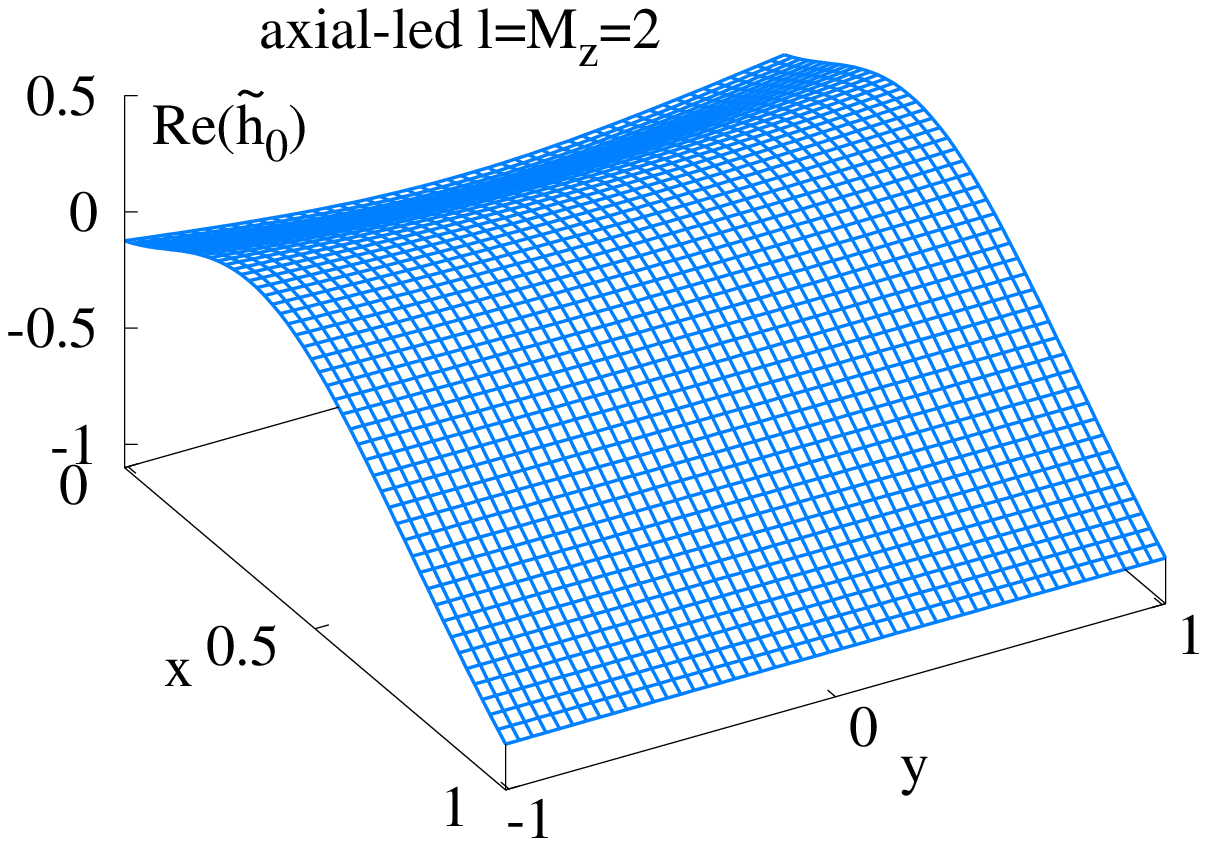}}

\subfloat[]{
\includegraphics[width=5.5cm,angle=-90]{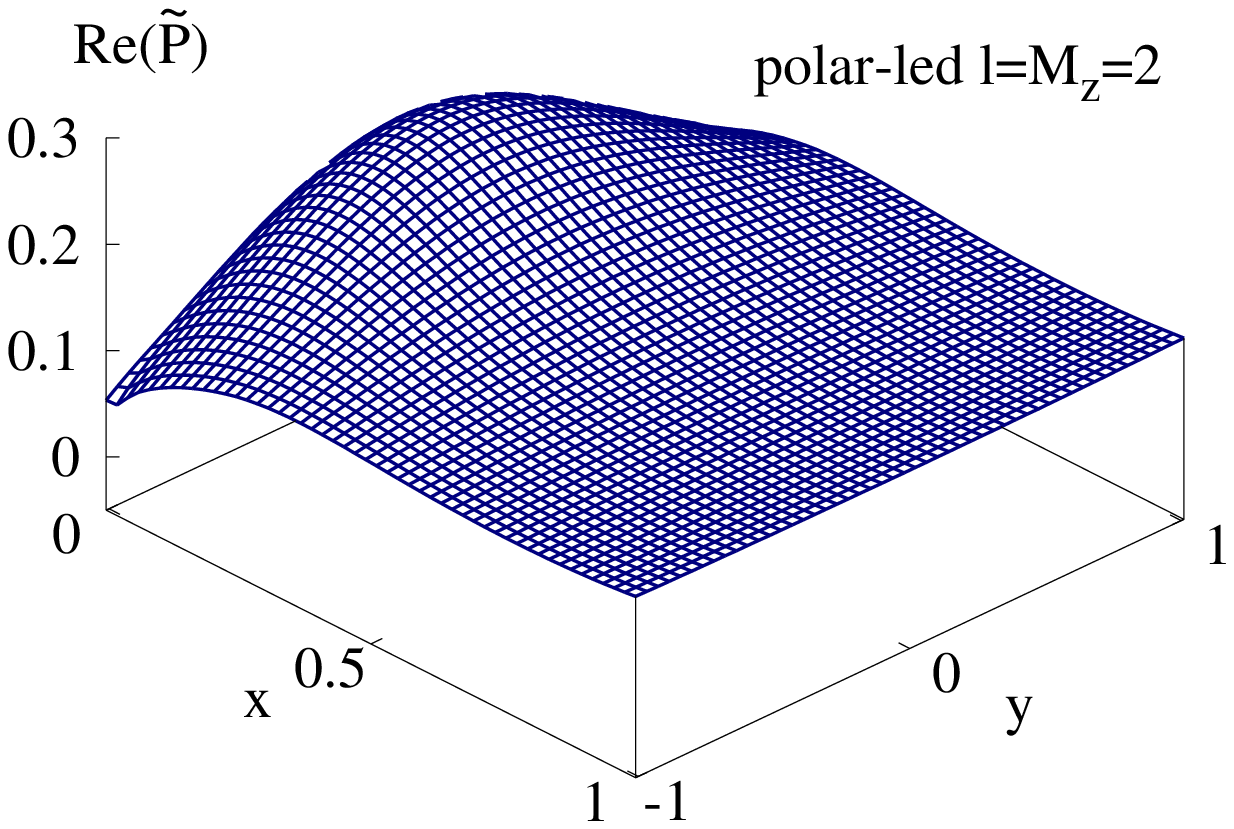}}
\subfloat[]{
\includegraphics[width=5.5cm,angle=-90]{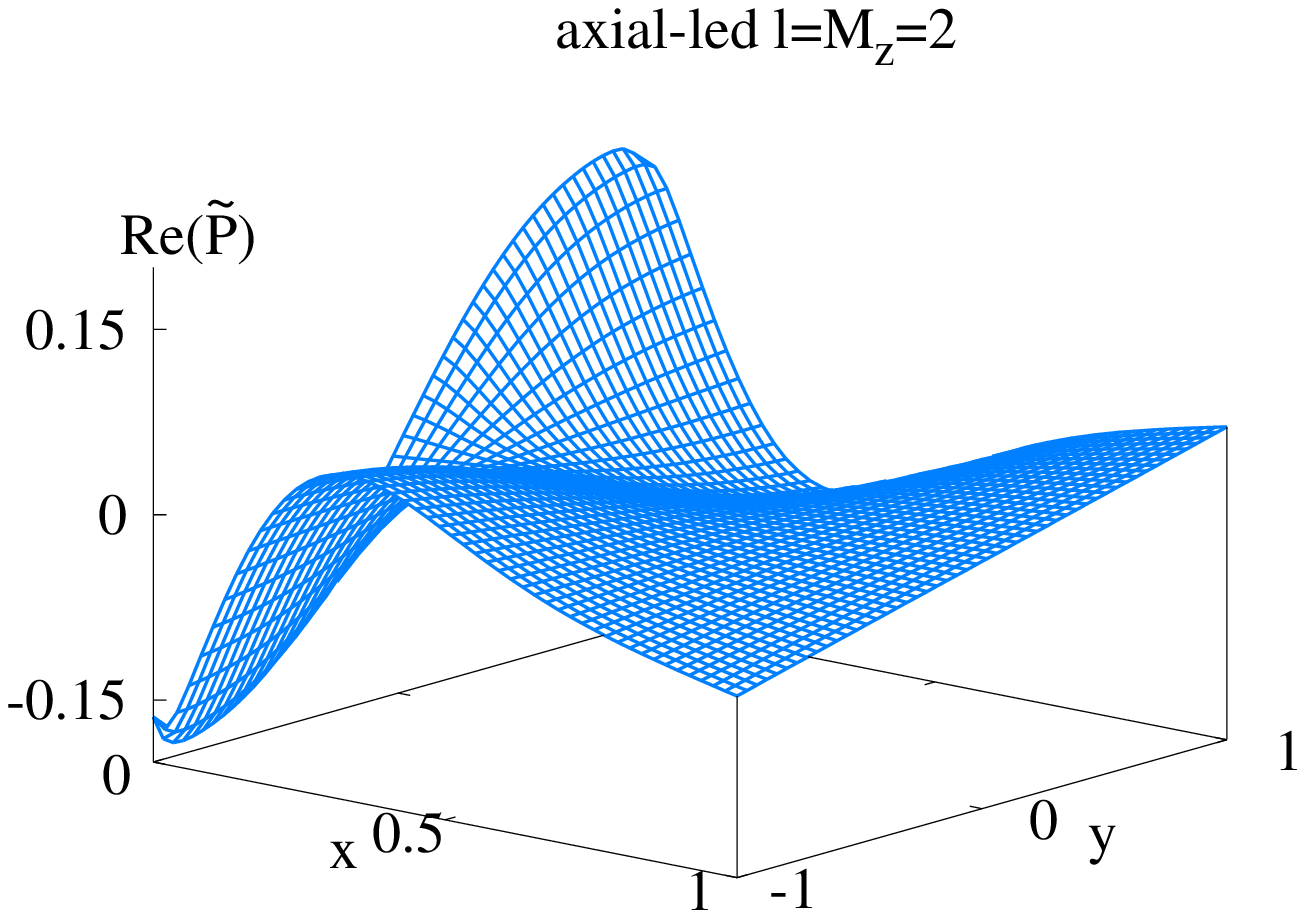}}
\end{center}
\caption{
Perturbation functions for $l=M_z=2$ fundamental modes for scaled angular momentum $j=0.6$ and scaled GB coupling $\xi=0.11$: 
function Re($\widetilde T$) for the polar-led mode (a) and the axial-led mode (b), function Re($\widetilde h_0$) for the polar-led mode (c) and the axial-led mode (d), function Re($\widetilde P$) for the polar-led mode (e) and the axial-led mode (f).
}
\label{fig_sol} 
\end{figure}

\begin{figure}[p!]
\begin{center}
\includegraphics[width=5.5cm,angle=-90]{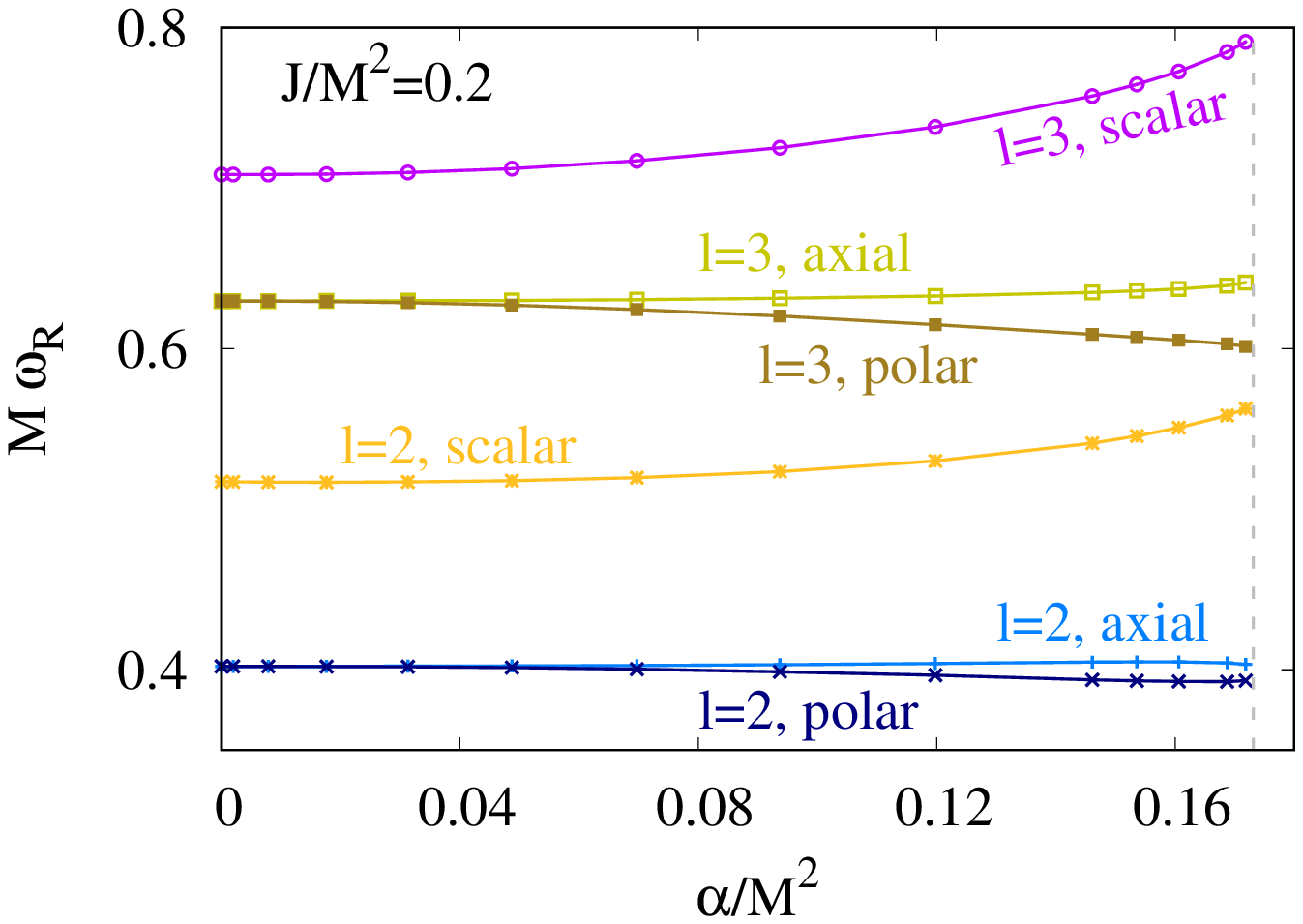}
\includegraphics[width=5.5cm,angle=-90]{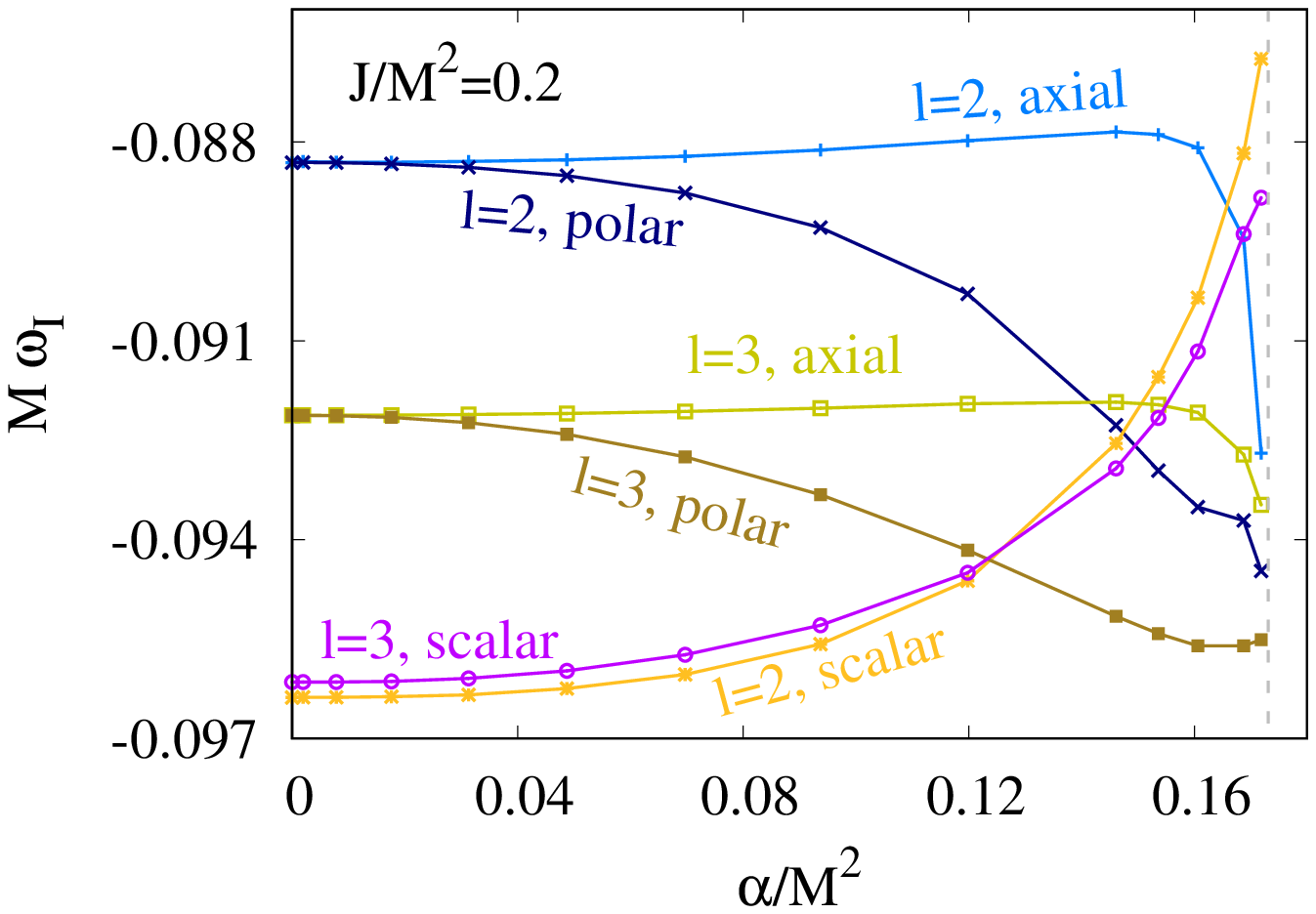}
\includegraphics[width=5.5cm,angle=-90]{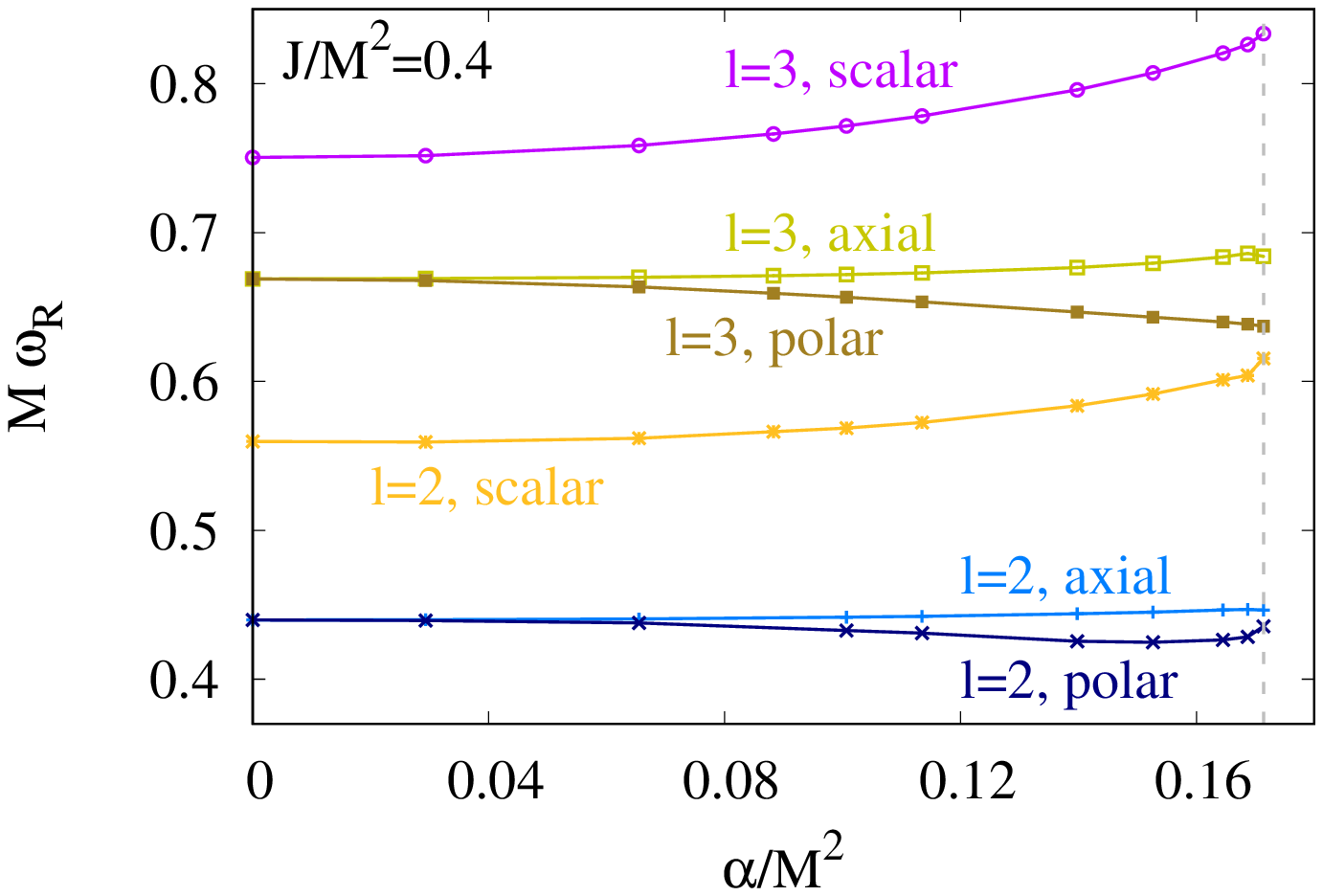}
\includegraphics[width=5.5cm,angle=-90]{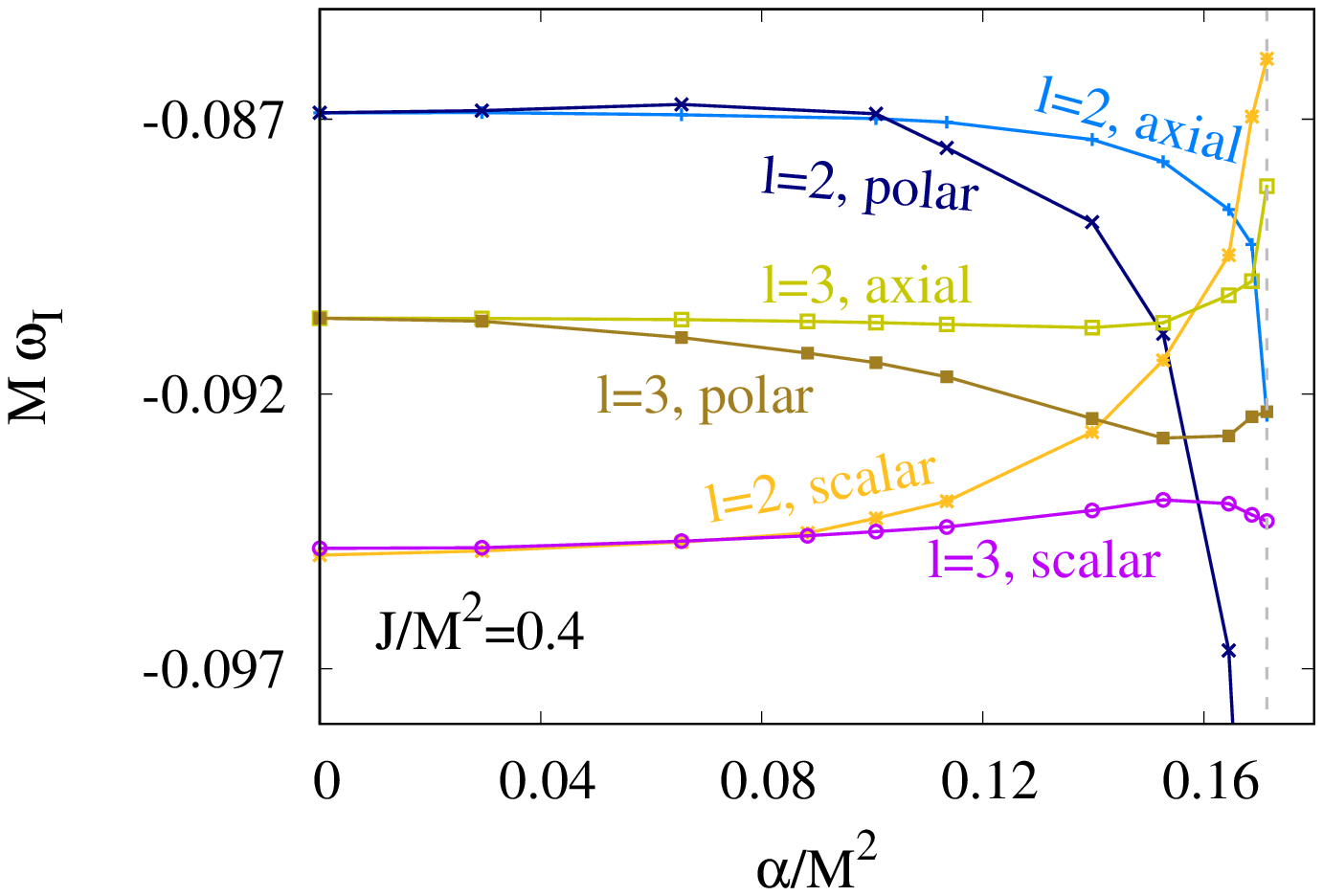}
\includegraphics[width=5.5cm,angle=-90]{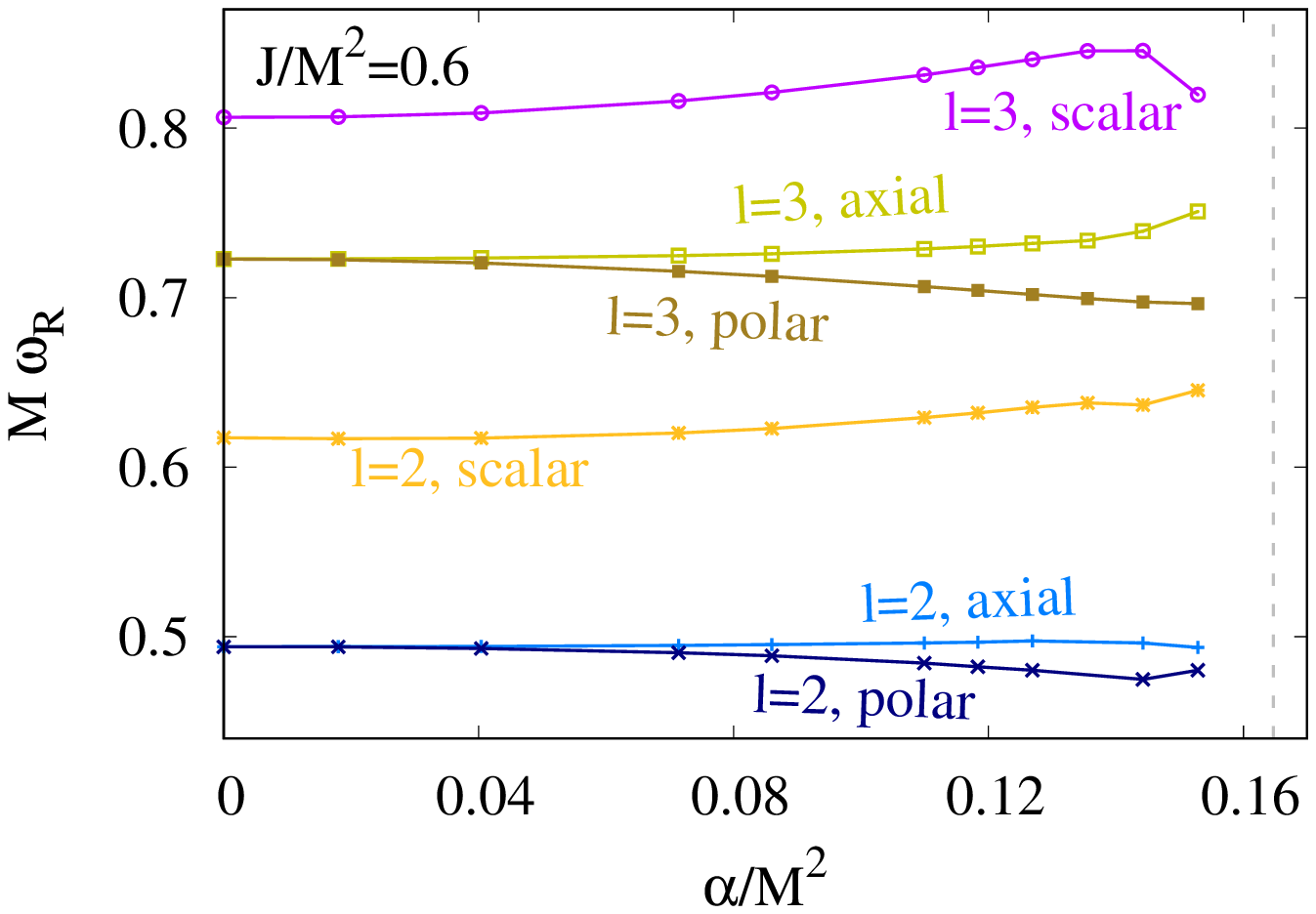}
\includegraphics[width=5.5cm,angle=-90]{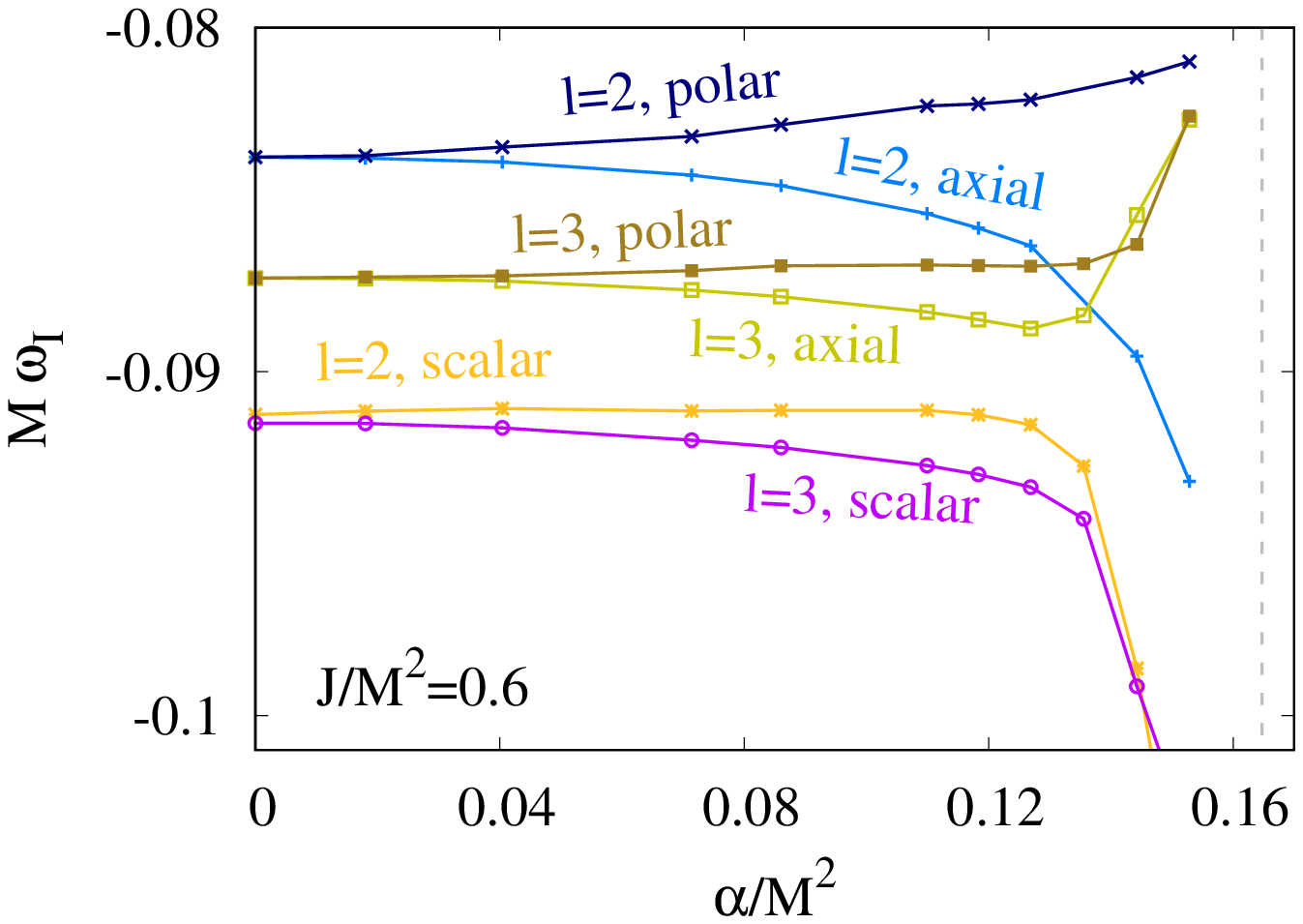}
\includegraphics[width=5.5cm,angle=-90]{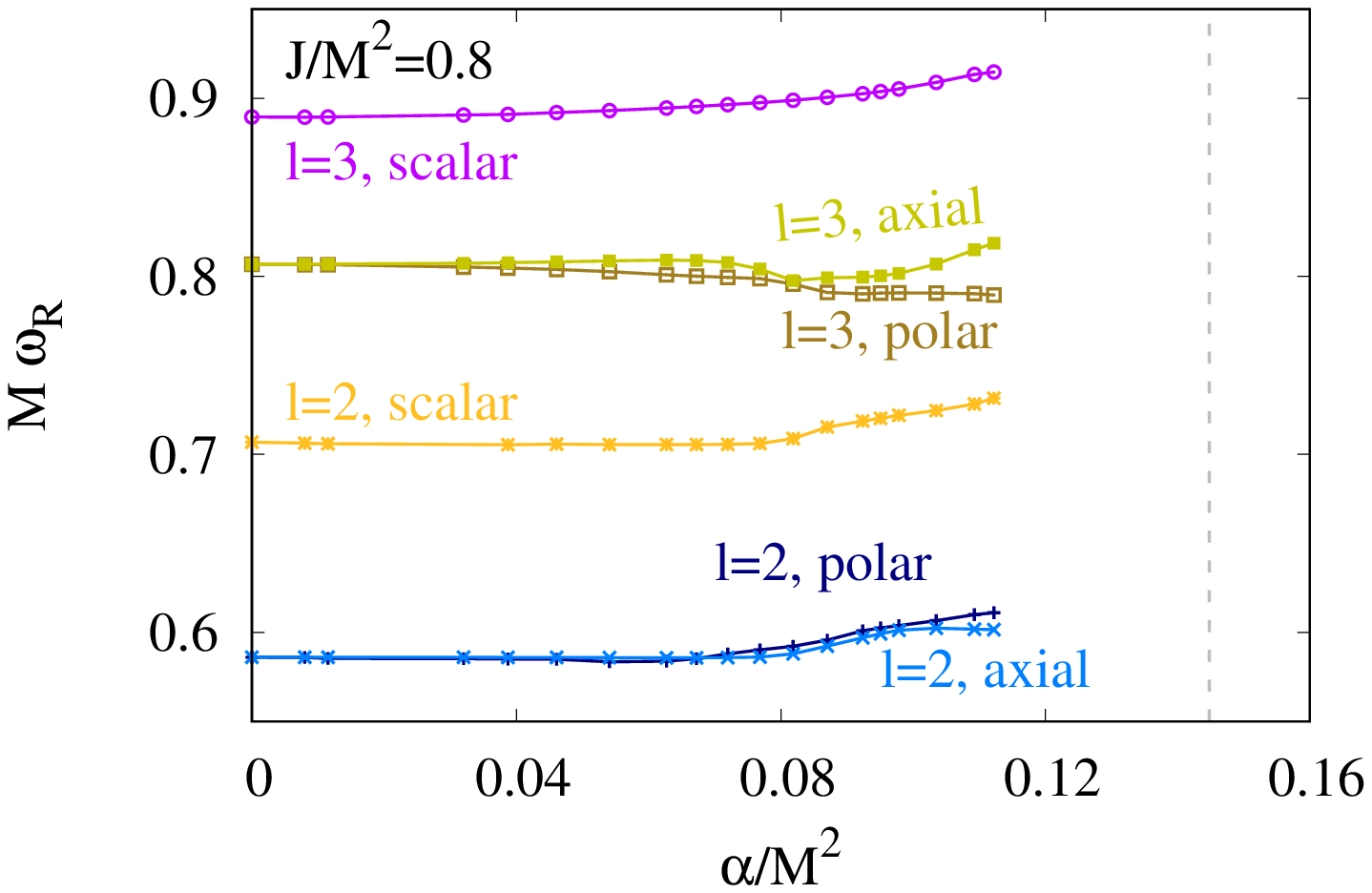}
\includegraphics[width=5.5cm,angle=-90]{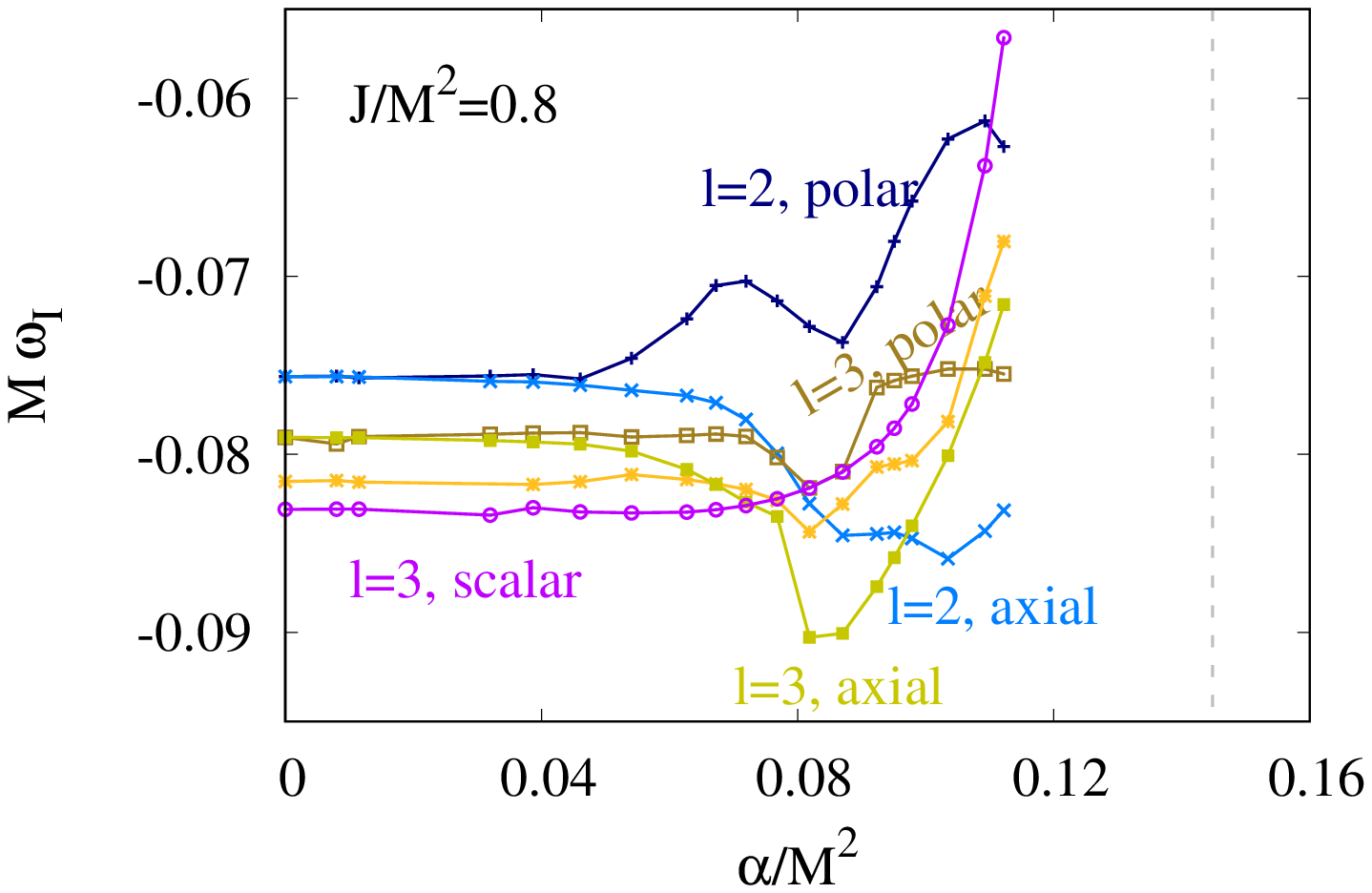}
\end{center}
\caption{Fundamental EGBd quadrupole $(l=2)$-led and octupole $(l=3)$-led quasinormal modes for $M_z=2$:
scaled real part $M \omega_R$ (left column) and scaled imaginary part $M \omega_I$ (right column) vs the scaled coupling strength $\xi$ for $j=0.2$, $j=0.4$, $j=0.6$, and $j=0.8$ (from top to bottom).
}
\label{fig_all_js} 
\end{figure}

In order to identify the type of a given mode, and classify it as polar-led, axial-led or scalar-led inspection of the perturbation functions is constructive.
We demonstrate this with the help of Fig.~\ref{fig_sol}, where we show the functions Re($\widetilde T$), Re($\widetilde h_0$), and Re($\widetilde P$) for the $l=M_z=2$ fundamental polar-led mode (left column) and the fundamental axial-led mode (right column) for scaled angular momentum $j=0.6$ and scaled GB coupling $\xi=0.11$.

Considering first the functions for the polar-led mode (left column) we note that Re($\widetilde T$) (a) and Re($\widetilde P$) (e) are predominantly 
even 
functions with respect to the $y$-coordinate, while Re($\widetilde h_0$) (c) behaves predominantly like an odd function.
For the axial-led mode (right column), on the other hand, this behavior is interchanged with Re($\widetilde T$) (b) and Re($\widetilde P$) (f) being predominantly odd functions, and  Re($\widetilde h_0$) (d) a predominantly even function.
We further note that for both modes the scalar function is non-negligible but sub-leading.

We now turn to the systematic analysis of the quasinormal {modes} as shown in Fig.~\ref{fig_all_js}.
Corresponding tables  (Table \ref{tab:j0.2-l2-grav} - Table \ref{tab:j0.8-l3-sc}) are found in the appendix.
The figure exhibits the dependence of the quasinormal modes of rotating EGBd black holes on the scaled coupling constant $\xi=\alpha/M^2$ for scaled angular momentum $j=0.2$, 0.4, 0.6, and 0.8.
Shown are the scaled real part $M\omega_R$ (left column) and the scaled imaginary part $M\omega_I$ (right column) for the quadrupole ($l=2$-led) and the octupole ($l=3$-led) modes.
For both multipoles the polar-led, axial-led and scalar-led modes are included.
The dotted vertical lines represent the maximal value of the scaled coupling constant, for which the corresponding background solutions exist (see Fig.~\ref{fig_bg}).

Isospectrality, which is present for the Kerr black holes at $\xi=0$, is clearly broken, when the {GB} coupling is turned on.
Typically the breaking of isospectrality and thus the splitting of the polar and axial modes increases toward the larger couplings.
Similarly the scalar modes typically deviate from the corresponding Kerr scalar test field mode the more, the larger the coupling is.
While the deviations of the real part $\omega_R$ do not exhibit very large changes with increasing $\xi$, this is often different for the imaginary parts $\omega_I$, where significant changes are seen toward the boundary of the domain of existence of rotating EGBd black holes.

For the frequencies $\omega_R$ of the modes we note that the order remains the same, starting from the $l=2$ polar modes up to the $l=3$ scalar modes, independent of the angular momentum $j$ and the coupling $\xi$.
The only exception is found for large $j$ and large $\xi$, where the $l=2$ polar and axial curves cross, making the axial modes the lowest modes beyond the crossing.
In contrast to the rather smooth dependencies of the frequencies on $j$ and $\xi$, the imaginary parts, and thus the damping times of the modes $\tau=1/|\omega_I|$, vary widely with $j$ and $\xi$.

For the small angular momentum $j=0.2$ the axial quadrupole-led mode is the dominant mode for most of the range of the coupling $\xi$.
However close to the boundary of the domain of existence there arises a sharp drop of $\omega_I$ of this previously dominating axial mode.
In fact, towards the boundary of the domain of existence both scalar modes cross all metric modes, with the $l=2$ scalar mode becoming finally the longest lived mode and the $l=3$ scalar mode becoming the second longest lived mode.
Thus the order of the modes and hence their relevance in the ringdown process is changing drastically.

As we increase the angular momentum, the order of the modes keeps changing towards the boundary of the domain of existence.
For $j=0.4$ the scalar quadrupole mode still becomes the dominant mode, but the scalar octupole mode remains short lived, and instead the axial octupole mode becomes the second longest lived mode.
For $j=0.6$ the polar quadrupole mode is always the dominant mode, while the polar and axial octupole modes become an almost degenerate pair of modes approaching the polar {quadrupole} mode.
Finally, for $j=0.8$ the scalar octupole mode becomes the leading mode for large coupling.
We note, however, that for $j=0.6$ and $0.8$ our calculations lose accuracy towards the boundary of the domain of existence.
We have therefore refrained from showing their values there.

The physical reason for the change of order of the modes and, in particular, the at first sight surprising dominance of the scalar modes toward the boundary of the domain of existence resides on the fact, that toward this boundary the scalar field of the background solutions increases considerably.
This then entails a stronger mixing of the metric and scalar contributions of the modes.
In any case, this change of order of the modes reveals that the dominant mode is only found by investigating all three type of modes, polar, axial and scalar, for the lowest fundamental multipoles.

\begin{figure}[t!]
\begin{center}
\includegraphics[width=5.5cm,angle=-90]{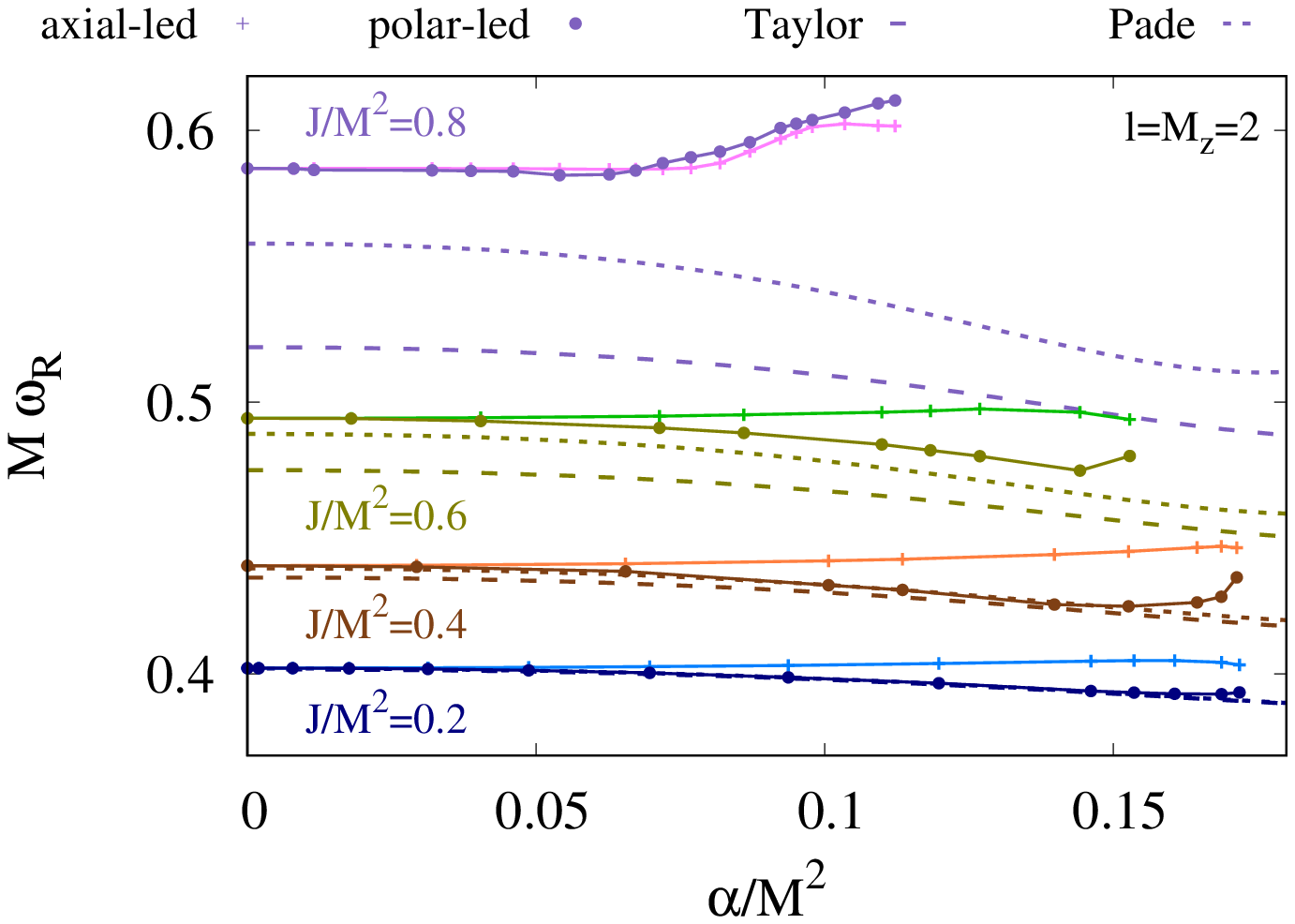}
\includegraphics[width=5.5cm,angle=-90]{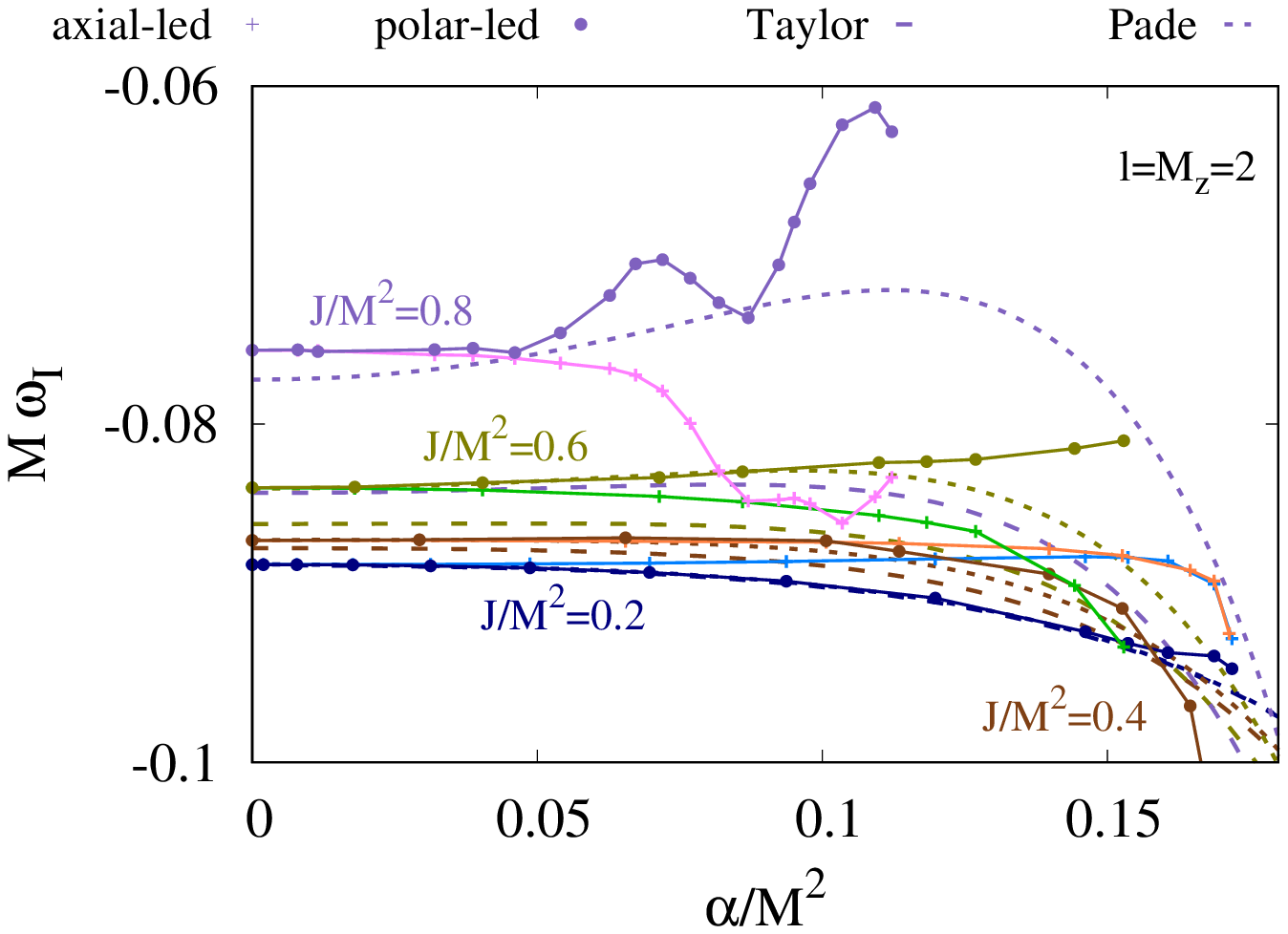}
\includegraphics[width=5.5cm,angle=-90]{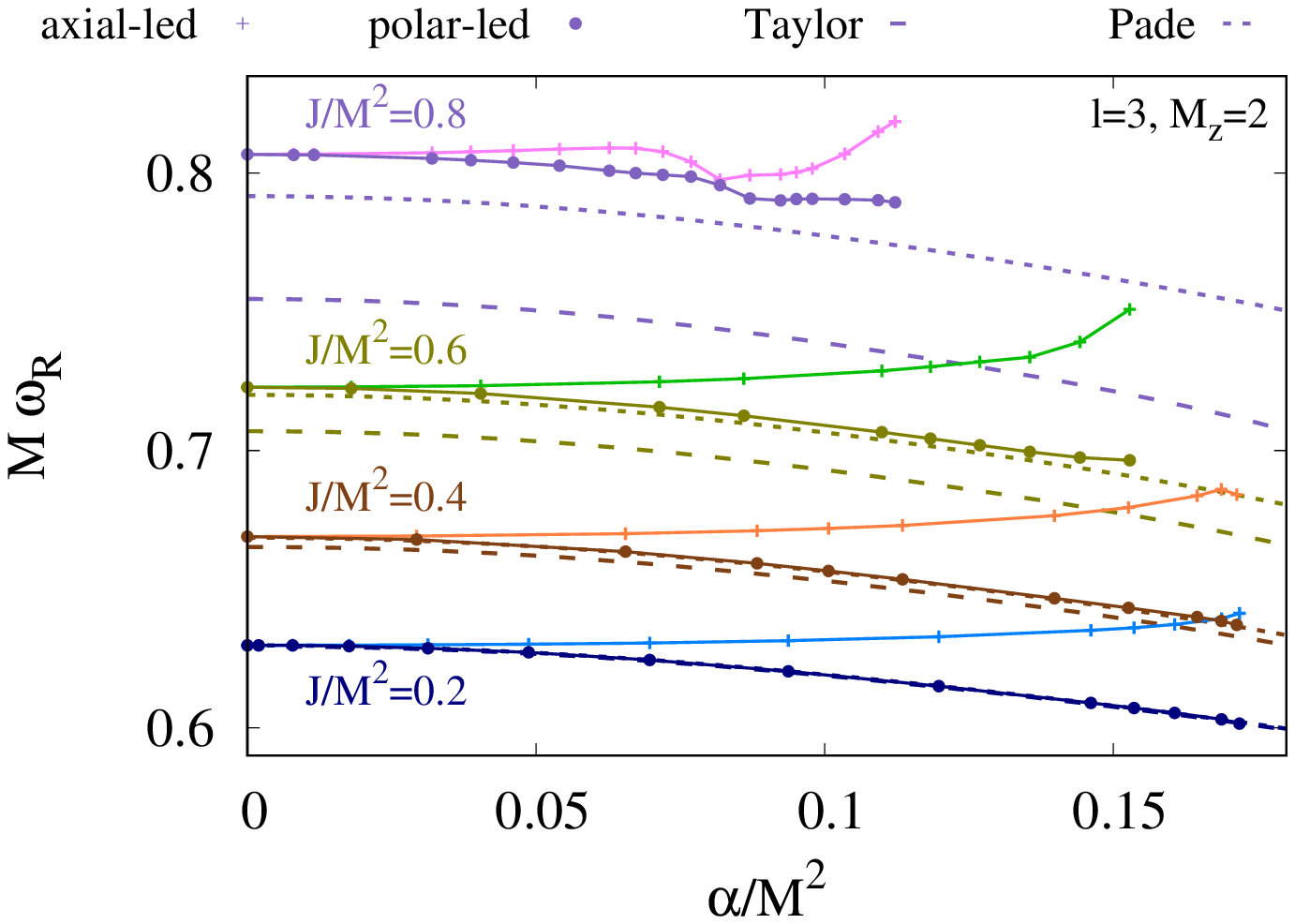}
\includegraphics[width=5.5cm,angle=-90]{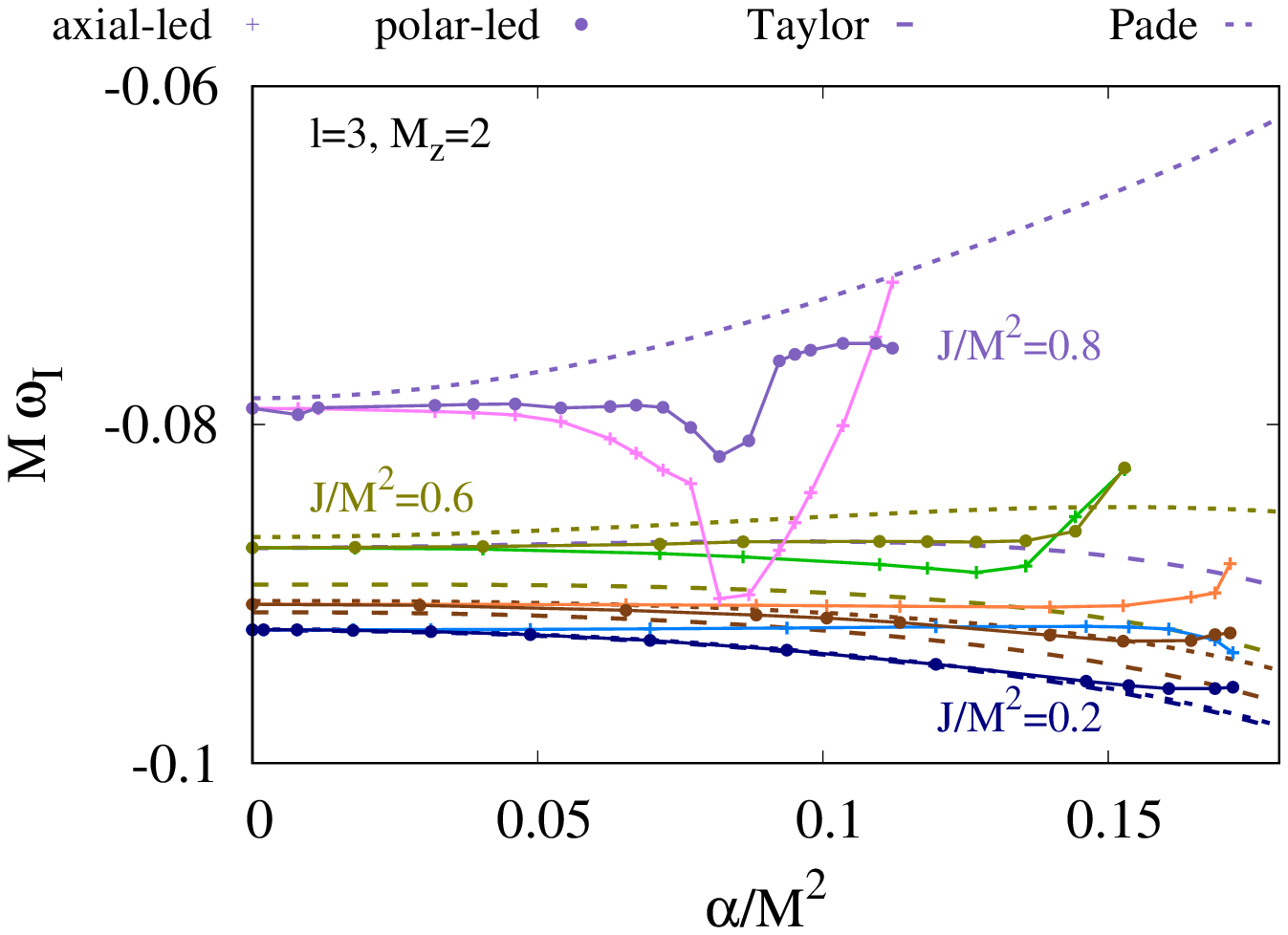}
\end{center}
\caption{Comparison with second order approximation: 
Fundamental EGBd $(l=2)$-led (upper row) and $(l=3)$-led (lower row) quasinormal modes for $M_z=2$:
scaled real part $M \omega_R$ (left column) and scaled imaginary part $M \omega_I$ (right column) vs scaled coupling strength $\xi$ for polar-led and axial-led exact modes (solid) and perturbative polar modes with Pade approximation (dotted) and Taylor expansion (dashed), for scaled angular momentum $j=0.2$, $j=0.4$, $j=0.6$, and $j=0.8$.
}
\label{fig_l2m2_vs_approx} 
\end{figure}

\begin{figure}[p!]
\begin{center}
\includegraphics[width=5.5cm,angle=-90]{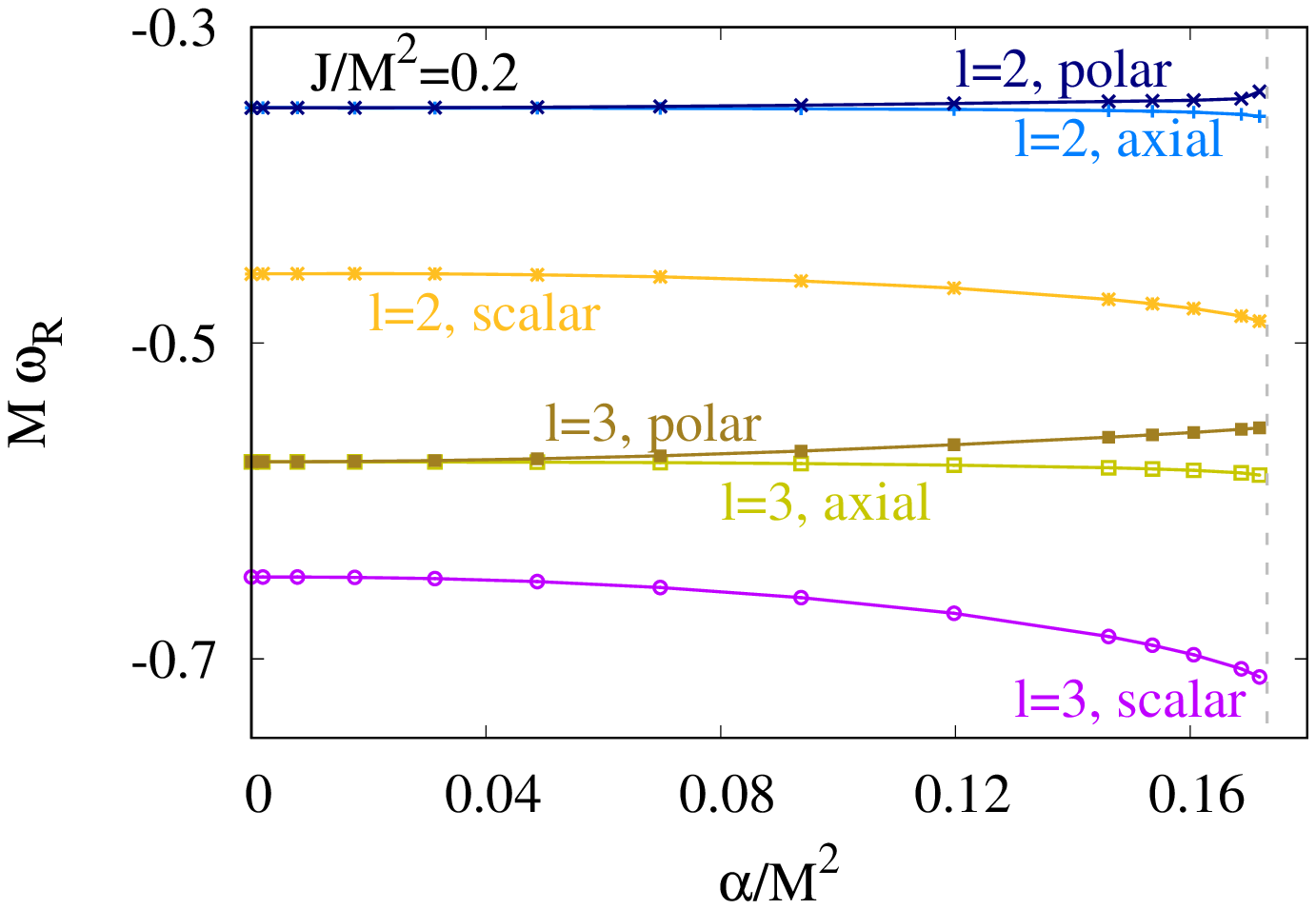}
\includegraphics[width=5.5cm,angle=-90]{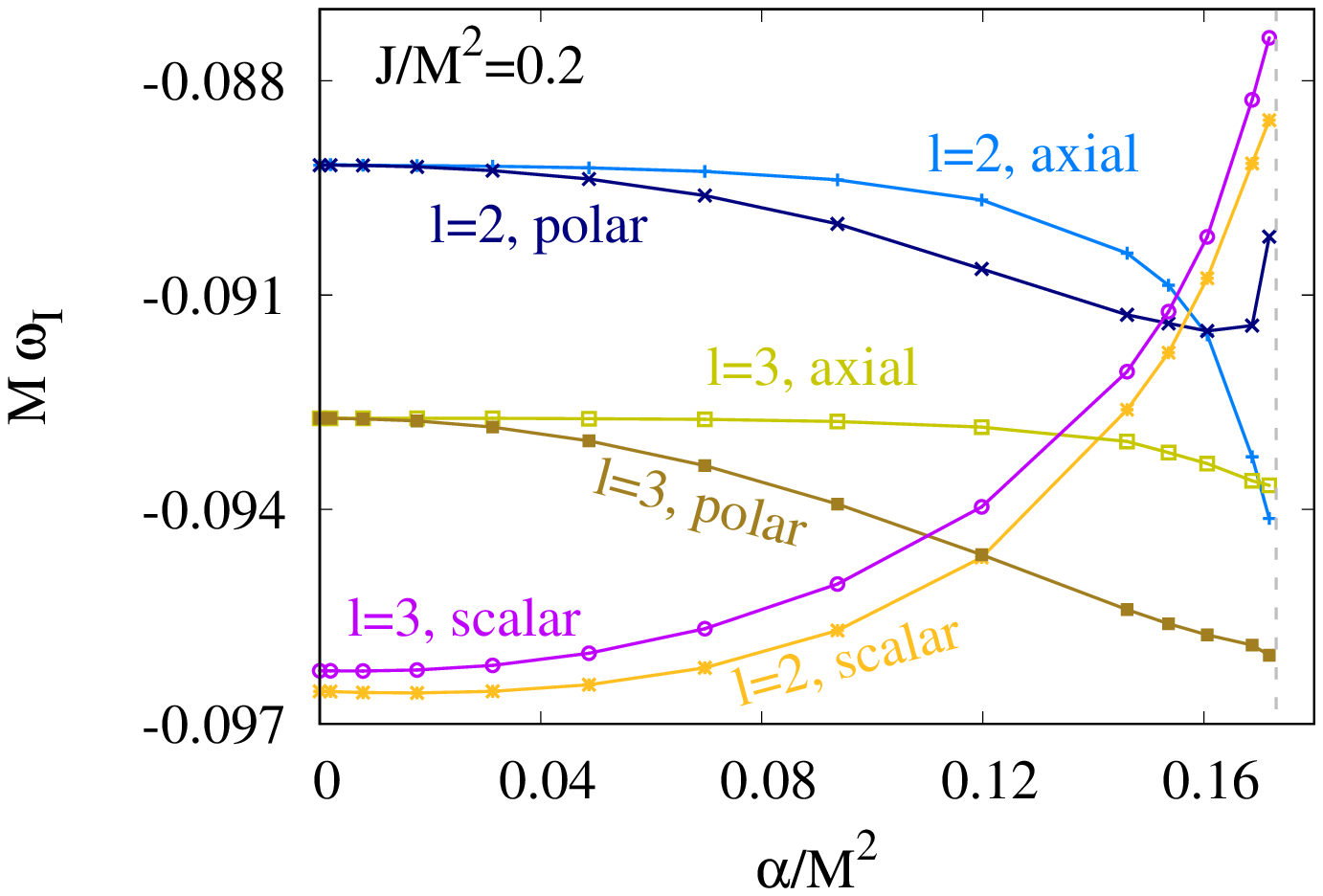}
\includegraphics[width=5.5cm,angle=-90]{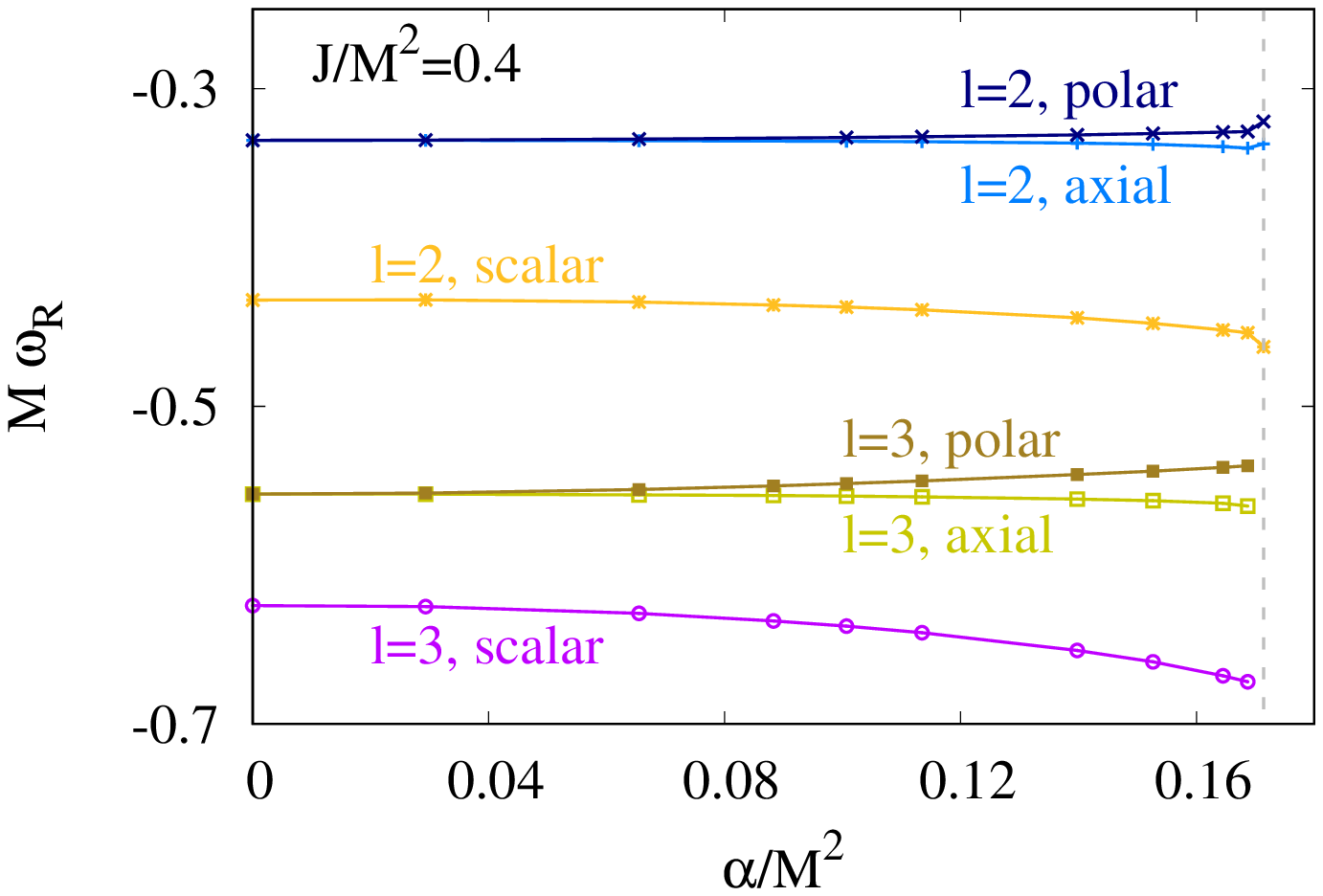}
\includegraphics[width=5.5cm,angle=-90]{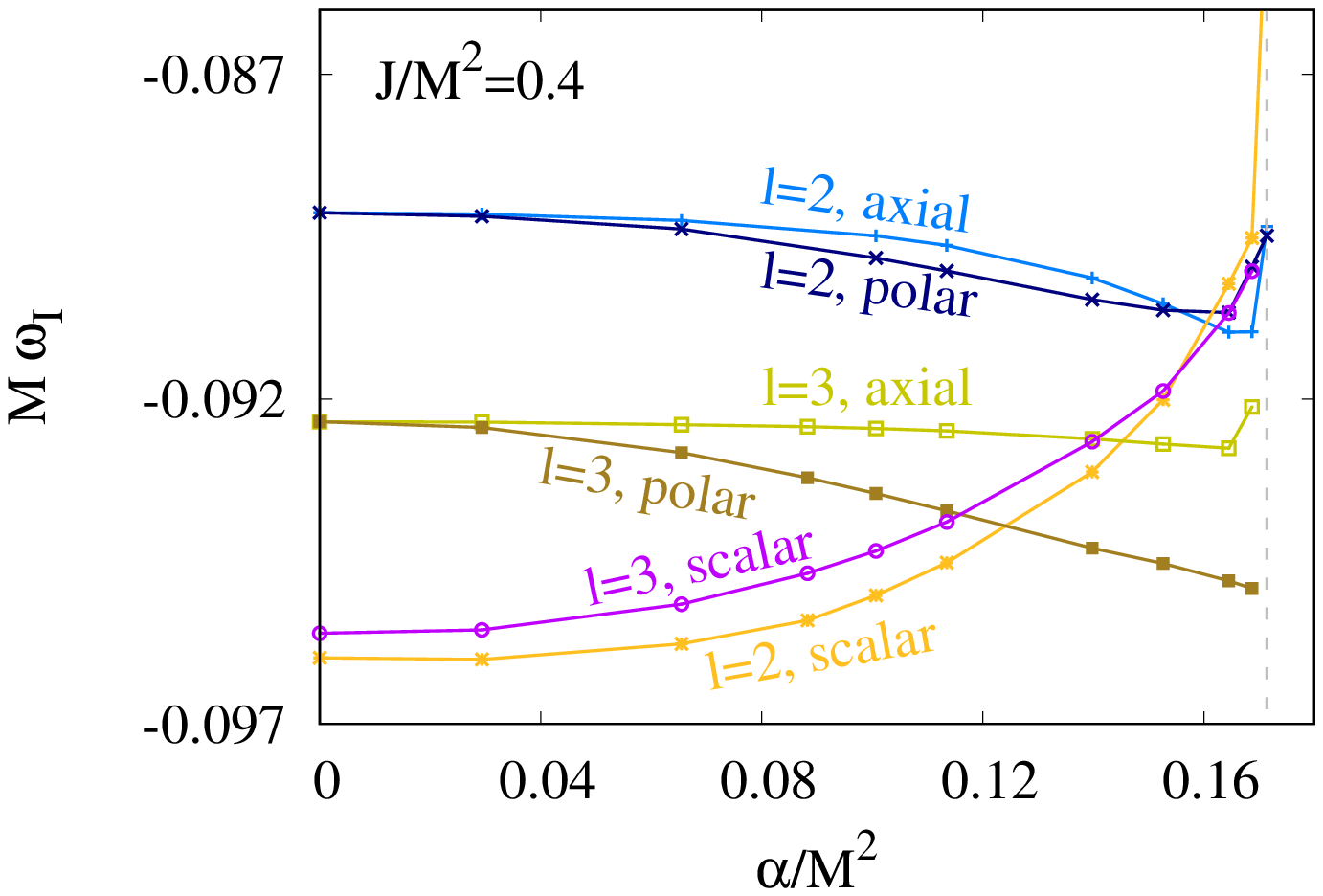}
\includegraphics[width=5.5cm,angle=-90]{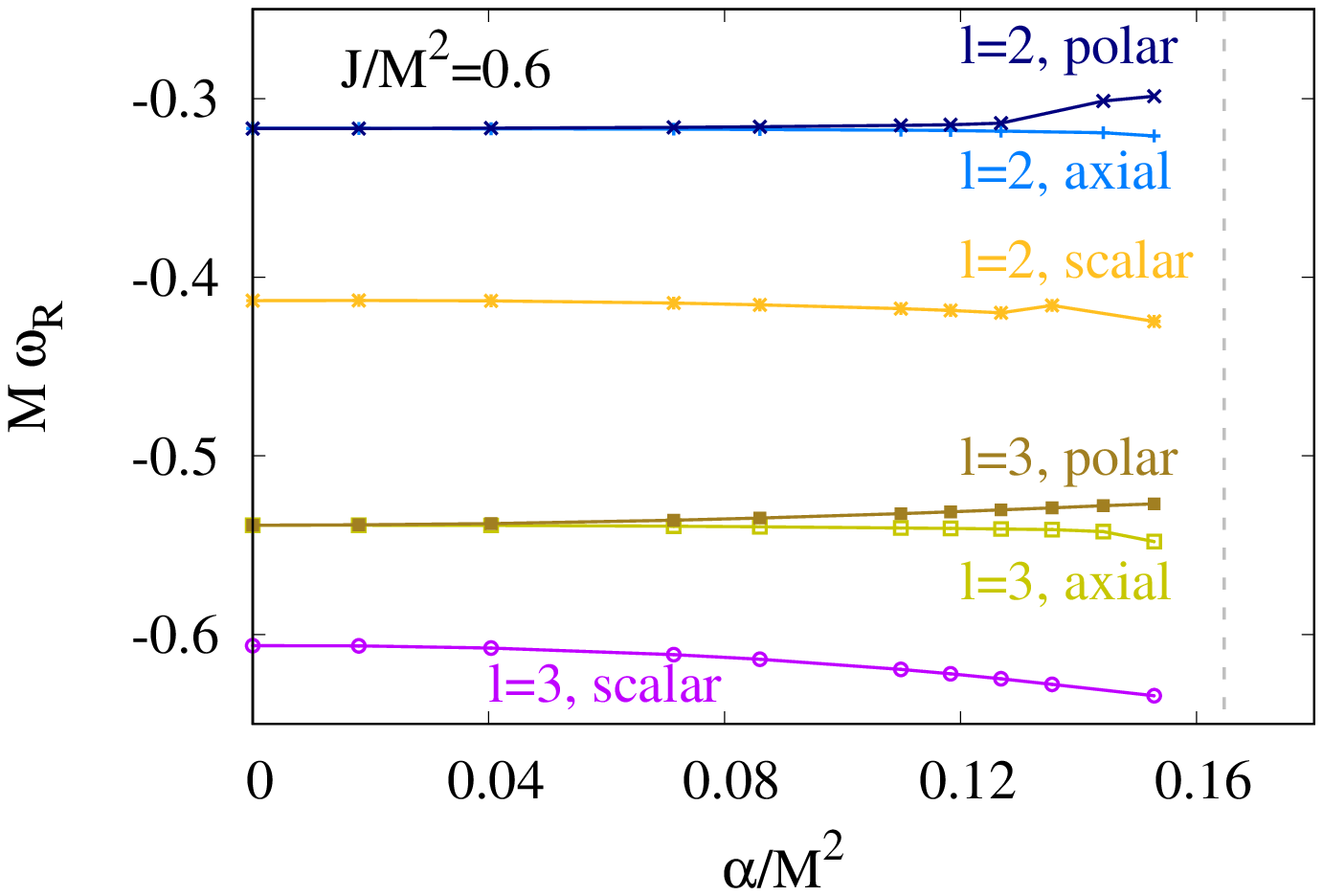}
\includegraphics[width=5.5cm,angle=-90]{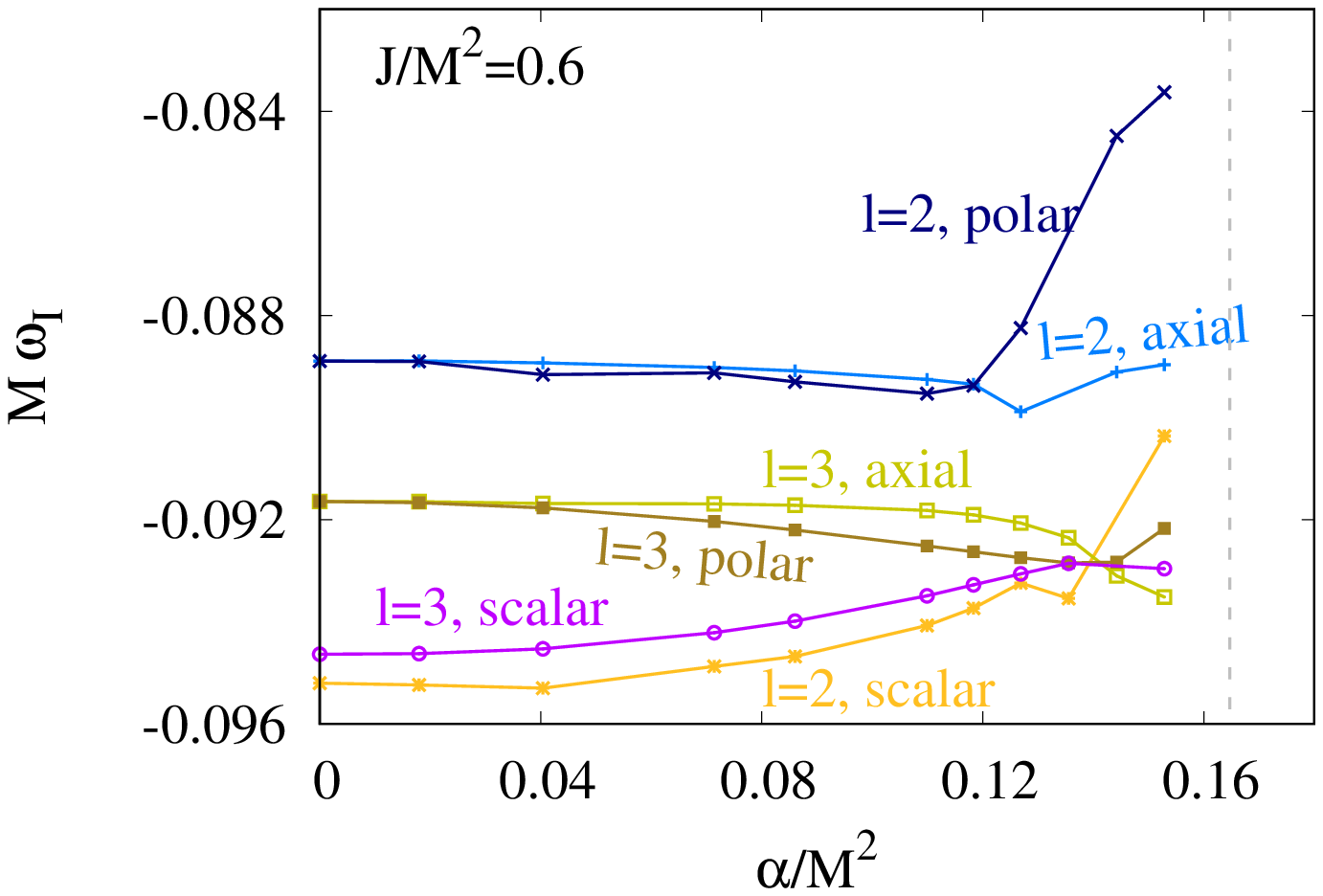}
\includegraphics[width=5.5cm,angle=-90]{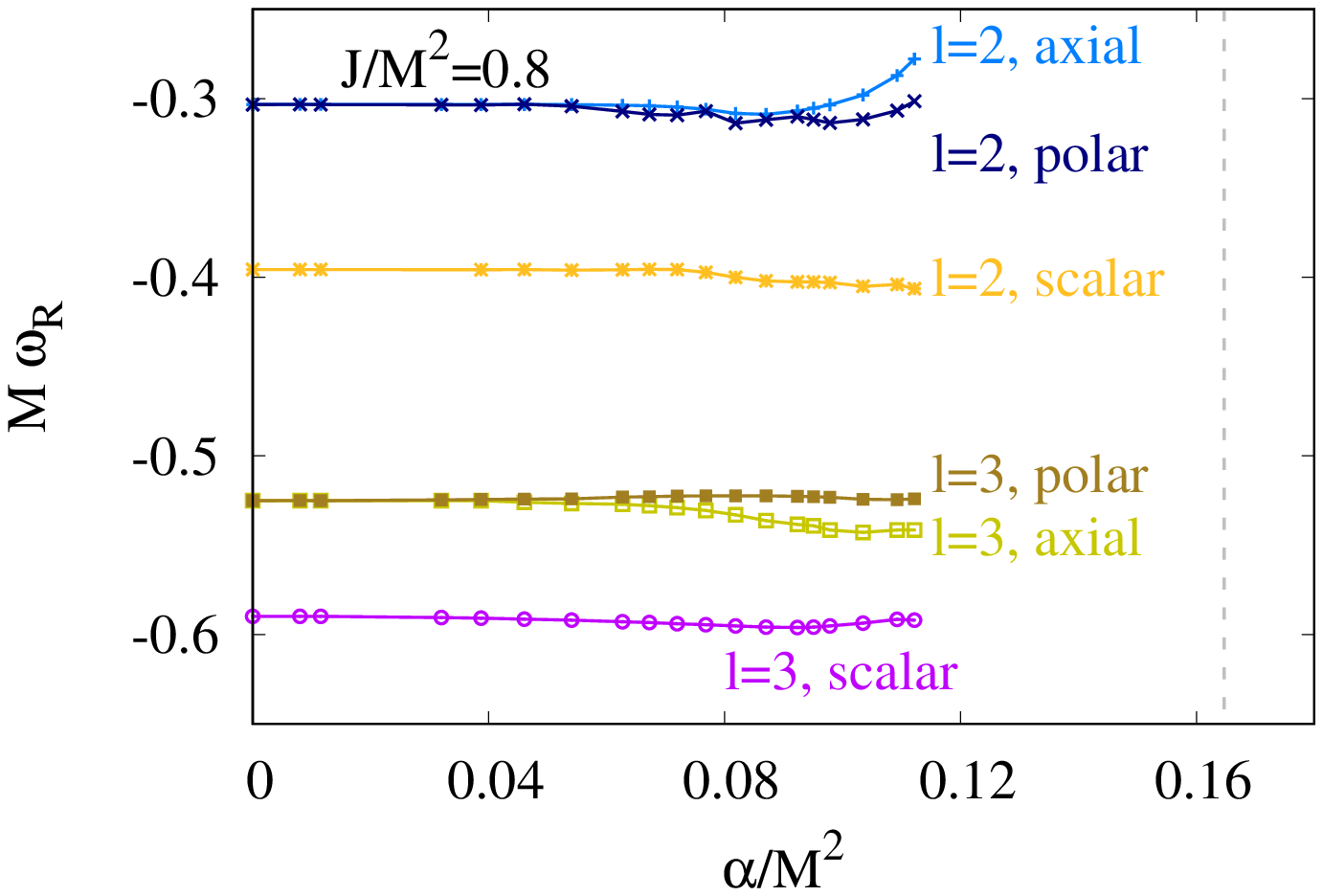}
\includegraphics[width=5.5cm,angle=-90]{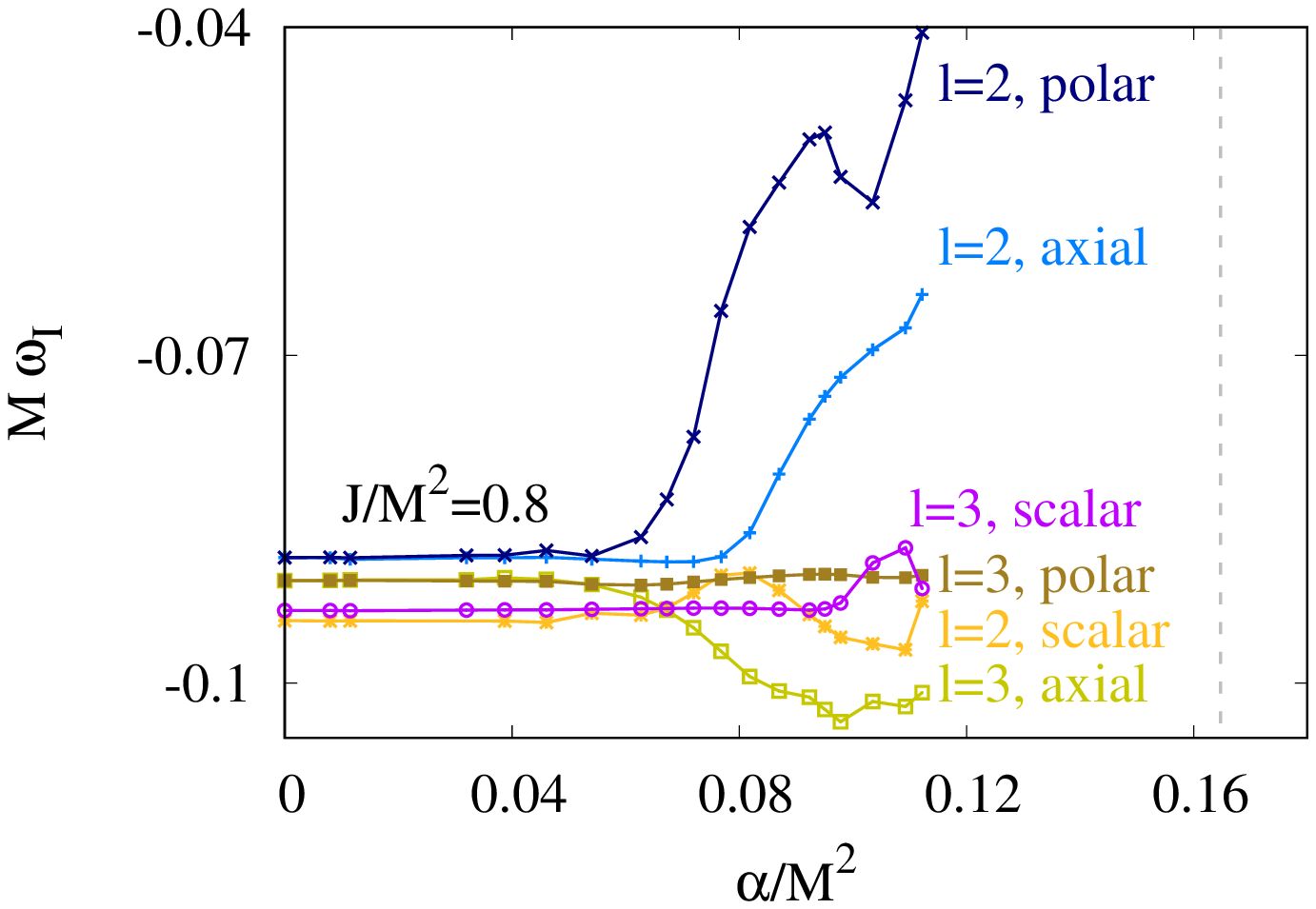}
\end{center}
\caption{Fundamental EGBd quadrupole $(l=2)$-led and octupole $(l=3)$-led quasinormal modes for $M_z=2$ for negative $\omega_R$:
scaled real part $M \omega_R$ (left column) and scaled imaginary part $M \omega_I$ (right column) vs the scaled coupling strength $\xi$ for $j=0.2$, $j=0.4$, $j=0.6$, and $j=0.8$ (from top to bottom).
}
\label{fig_all_js_-} 
\end{figure} 

It is instructive to compare our exact results with previously obtained perturbative results, that employ perturbation theory both for the angular momentum $j$ and for the coupling $\xi$, going to second order in $j$ and to $6$th order in $\xi$ \cite{Pierini:2022eim}.
We exhibit this comparison in Fig.~\ref{fig_l2m2_vs_approx}, where we show our axial and polar quadrupole modes (upper row) and octupole modes (lower row).
Recall, that the domain of existence of the modes shrinks with increasing angular momentum.

In addition, the figure includes the polar modes as obtained in perturbation theory \cite{Pierini:2022eim}, while the axial modes were not given in there.
These modes as obtained from the Taylor expansion are shown by dashed curves of the same color as the associated exact modes (solid).
We observe good agreement for the small angular momenta, but increasingly strong deviations for the larger angular momenta, as present already for the Kerr case, since only second order perturbation theory in $j$ is employed \cite{Pierini:2022eim}.
If we make use of a Pad\'e approximation instead (dotted), the agreement improves substantially.
In particular, for $j=0.2$ the agreement is excellent almost up to the boundary of the domain of existence, and for $j=0.4$ the comparison still reveals an amazingly good agreement for the frequency $\omega_R$.

We finally address the quasinormal modes for negative frequencies $\omega_R$.
These are depicted in Fig.~\ref{fig_all_js_-} analogously to Fig.~\ref{fig_all_js} where the positive frequencies are shown.
For the negative frequencies we observe quite similar smooth behavior as for the positive frequencies, i.e., the absolute values exhibit the same order of the modes, with the polar quadrupole mode possessing the smallest $|\omega_R|$ except for $j=0.8$ and large coupling $\xi$.
Similarly to the positive frequency modes, the damping times of the negative frequency modes show strong dependencies on the angular momentum and the coupling.
For $j=0.2$ the scalar $l=2$ and $l=3$ {modes} become the longest lived modes for large coupling, but now $l=3$ {mode} becomes the dominant one.
For $j=0.4$ the scalar $l=2$ mode becomes the dominant one here as well,
for $j=0.6$ the polar and axial $l=2$ are very close until toward the boundary of the domain of existence the polar mode dominates. 
However, for $j=0.8$ the polar $l=2$ dominates for negative modes and large couplings, instead of the scalar $l=3$ mode as in the positive frequency case.

\section{Conclusions}

We have presented a detailed derivation and discussion of the quasinormal modes of rotating EGBd black holes, where neither the GB coupling nor the rotation was treated in a perturbative way.
Our starting point is therefore the construction of an appropriate set of rotating EGBd black holes, that serve as the background solutions for the perturbations.
This procedure also clarifies the domain of existence of rotating black holes, if not yet known, and thus indicates where perturbative procedures would break down.

We have focused on the calculation of polar, axial and scalar $l=2$-led and $l=3$-led modes for $M_z=2$, for both positive and negative frequencies.
Our calculations reveal, that the order of the modes can change with increasing coupling strength, and thus the dominant mode in the ringdown would depend on the GB coupling of the theory and the angular momentum of the final black hole.
In particular, toward the boundary of the domain of existence of background solutions, the scalar modes tend to dominate, due to the large background scalar field.
Moreover, the octupole modes may also dominate over the quadrupole modes for large coupling.

Consequently, the gravitational radiation emitted in the ringdown may differ strongly for EGBd theory (and possibly other alternative theories of gravity) from the radiation in GR.
Future gravitational wave observations will then allow to put bounds on the coupling strength from precision measurements of the ringdown.
When theories possess a scalar degree of freedom, however, the resulting dipole radiation during the inspiral phase can already now put new bounds \cite{Julie:2024fwy}.

Currently we are studying the exact quasinormal modes of shift-symmetric EGB theory, which has coupling function $f(\varphi)=\varphi$.
The static and rotating black holes of this theory were constructed before \cite{Sotiriou:2013qea,Sotiriou:2014pfa,Delgado:2020rev,Sullivan:2020zpf}.
Recently the quasinormal modes of shift-symmetric black holes were obtained for rapid rotation, but only in second order perturbation theory in the coupling constant \cite{Chung:2024ira,Chung:2024vaf}.
Thus these results are no longer appropriate for the larger couplings, in particular, close to the boundary of the domain of existence of the background solutions.
Our calculations indicate the occurrence of rather large deviations 
of the perturbative results from the exact ones in this case \cite{Khoo:2024agm}.

Future studies will address the overtones of the modes of rotating black holes in EGBd theory.
They will address as well the modes of rotating spontaneously scalarized black holes in Einstein-Gau\ss-Bonnet-scalar theories, including the spin-induced black holes.
Indeed, the machinery can be applied to the rotating black holes of all kinds of Horndeski theories.
Moreover, the quasinormal modes of other types of compact objects can be tackled with these methods, like rapidly rotating neutron stars, boson stars and wormholes.

\section*{Acknowledgement}

We gratefully acknowledge support by DFG project Ku612/18-1, 
FCT project PTDC/FIS-AST/3041/2020, 
and MICINN project PID2021-125617NB-I00 ``QuasiMode''. JLBS gratefully acknowledges support from MICINN project CNS2023-144089 ``Quasinormal modes''.
FSK gratefully acknowledges support from ``Atracci\'on de Talento Investigador Cesar Nombela'' of the Comunidad de Madrid, grant no. 2024-T1/COM-31385.
Part of the computations were performed on the HPC Cluster ROSA at the University of Oldenburg,
funded by the DFG 
and the Ministry of Science and Culture (MWK) of the Lower Saxony State
under the grant number INST 184/225-1 FUGG.

\section*{Appendix}

Here we present a set of tables containing the numerical values of the fundamental  $(l=2)$-led and $(l=3)$-led quasinormal modes for the 
gravitational sector (polar, axial) and scalar sector for $M_z=2$.
For each angular momentum $j=0.2,0.4,0.6,0.8$,
we provide both the positive and negative real frequencies
$\omega_R$ of the modes, and their corresponding imaginary parts $\omega_I$ in the tables.
The EGBd modes are computed for values of the scaled coupling constant $\xi$ up to the vicinity of its limit bounded by each $j$. 
The first row of the tables
gives Kerr quasinormal modes
where $\xi=0$.

\begin{table}
    \centering
    \begin{tabular}{|c||c|c||c|c||c|c||c|c|}
\hline
  \multirow{2}{*}{}  &
  \multicolumn{2}{|c||}{Axial} &
  \multicolumn{2}{|c||}{Polar} &
  \multicolumn{2}{|c||}{Axial} &
  \multicolumn{2}{|c|}{Polar} \\ 
  $\alpha/M^2$   & $\omega_R$ & $\omega_I$ & $\omega_R$ & $\omega_I$ & $\omega_R$ & $\omega_I$ & $\omega_R$ & $\omega_I$ \\
 \hline
0	&	0.40215	&	-0.08831	&	0.40215	&	-0.08831	&	-0.35105	&	-0.08918	&	-0.35105	&	-0.08918	\\
0.00196	&	0.40217	&	-0.08830	&	0.40215	&	-0.08831	&	-0.35105	&	-0.08917	&	-0.35105	&	-0.08918	\\
0.00784	&	0.40215	&	-0.08830	&	0.40213	&	-0.08831	&	-0.35105	&	-0.08918	&	-0.35104	&	-0.08919	\\
0.01763	&	0.40218	&	-0.08830	&	0.40205	&	-0.08833	&	-0.35107	&	-0.08919	&	-0.35100	&	-0.08921	\\
0.03129	&	0.40224	&	-0.08829	&	0.40182	&	-0.08838	&	-0.35110	&	-0.08920	&	-0.35087	&	-0.08926	\\
0.04872	&	0.40238	&	-0.08827	&	0.40132	&	-0.08851	&	-0.35118	&	-0.08922	&	-0.35061	&	-0.08937	\\
0.06969	&	0.40265	&	-0.08822	&	0.40037	&	-0.08877	&	-0.35132	&	-0.08927	&	-0.35013	&	-0.08961	\\
0.09370	&	0.40312	&	-0.08812	&	0.39877	&	-0.08929	&	-0.35157	&	-0.08939	&	-0.34936	&	-0.09000	\\
0.11979	&	0.40383	&	-0.08798	&	0.39656	&	-0.09029	&	-0.35199	&	-0.08967	&	-0.34825	&	-0.09063	\\
0.14613	&	0.40471	&	-0.08785	&	0.39374	&	-0.09228	&	-0.35274	&	-0.09041	&	-0.34692	&	-0.09128	\\
0.15360	&	0.40490	&	-0.08789	&	0.39311	&	-0.09296	&	-0.35312	&	-0.09086	&	-0.34655	&	-0.09139	\\
0.16061	&	0.40493	&	-0.08809	&	0.39269	&	-0.09351	&	-0.35370	&	-0.09155	&	-0.34622	&	-0.09150	\\
0.16873	&	0.40424	&	-0.08945	&	0.39254	&	-0.09371	&	-0.35513	&	-0.09326	&	-0.34524	&	-0.09143	\\
0.17185	&	0.40334	&	-0.09270	&	0.39323	&	-0.09447	&	-0.35644	&	-0.09413	&	-0.34070	&	-0.09018	\\
 \hline

    \end{tabular}
    \caption{Gravitational $(l=2)$-led fundamental quasinormal modes with $M_z=2$ for different scaled coupling strength $\xi$ for $j=0.2$.}
    \label{tab:j0.2-l2-grav}
\end{table}

\begin{table}
    \centering
    \begin{tabular}{|c||c|c||c|c||c|c||c|c|}
\hline
  \multirow{2}{*}{}  &
  \multicolumn{2}{|c||}{Axial} &
  \multicolumn{2}{|c||}{Polar} &
  \multicolumn{2}{|c||}{Axial} &
  \multicolumn{2}{|c|}{Polar} \\ 
  $\alpha/M^2$   & $\omega_R$ & $\omega_I$ & $\omega_R$ & $\omega_I$ & $\omega_R$ & $\omega_I$ & $\omega_R$ & $\omega_I$ \\
 \hline
0.00000	&	0.62974	&	-0.09212	&	0.62974	&	-0.09212	&	-0.57517	&	-0.09272	&	-0.57517	&	-0.09272	\\
0.00196	&	0.62974	&	-0.09212	&	0.62973	&	-0.09212	&	-0.57517	&	-0.09272	&	-0.57517	&	-0.09272	\\
0.00784	&	0.62975	&	-0.09212	&	0.62967	&	-0.09213	&	-0.57517	&	-0.09272	&	-0.57512	&	-0.09273	\\
0.01763	&	0.62979	&	-0.09212	&	0.62941	&	-0.09216	&	-0.57520	&	-0.09272	&	-0.57493	&	-0.09276	\\
0.03129	&	0.62989	&	-0.09211	&	0.62868	&	-0.09223	&	-0.57527	&	-0.09273	&	-0.57441	&	-0.09285	\\
0.04872	&	0.63013	&	-0.09210	&	0.62715	&	-0.09241	&	-0.57543	&	-0.09273	&	-0.57330	&	-0.09304	\\
0.06969	&	0.63059	&	-0.09207	&	0.62445	&	-0.09275	&	-0.57574	&	-0.09274	&	-0.57135	&	-0.09339	\\
0.09370	&	0.63141	&	-0.09202	&	0.62037	&	-0.09332	&	-0.57629	&	-0.09277	&	-0.56836	&	-0.09392	\\
0.11979	&	0.63280	&	-0.09195	&	0.61500	&	-0.09416	&	-0.57725	&	-0.09285	&	-0.56436	&	-0.09463	\\
0.14613	&	0.63511	&	-0.09193	&	0.60892	&	-0.09516	&	-0.57891	&	-0.09305	&	-0.55962	&	-0.09540	\\
0.15360	&	0.63607	&	-0.09196	&	0.60711	&	-0.09542	&	-0.57965	&	-0.09320	&	-0.55811	&	-0.09560	\\
0.16061	&	0.63722	&	-0.09208	&	0.60535	&	-0.09560	&	-0.58056	&	-0.09336	&	-0.55659	&	-0.09576	\\
0.16873	&	0.63933	&	-0.09272	&	0.60301	&	-0.09560	&	-0.58227	&	-0.09360	&	-0.55453	&	-0.09590	\\
0.17185	&	0.64127	&	-0.09348	&	0.60145	&	-0.09551	&	-0.58365	&	-0.09366	&	-0.55367	&	-0.09604	\\
 \hline

    \end{tabular}
    \caption{Gravitational $(l=3)$-led fundamental quasinormal modes with $M_z=2$ for different scaled coupling strength $\xi$ for $j=0.2$.}
    \label{tab:j0.2-l3-grav}
\end{table}

\begin{table}
    \centering
    \begin{tabular}{|c||c|c||c|c|} 
    \hline
  $\alpha/M^2$   & $\omega_R$ & $\omega_I$ & $\omega_R$ & $\omega_I$  \\
 \hline
0.00000	&	0.51712	&	-0.09638	&	-0.45620	&	-0.09655	\\
0.00196	&	0.51706	&	-0.09638	&	-0.45616	&	-0.09655	\\
0.00784	&	0.51692	&	-0.09638	&	-0.45607	&	-0.09656	\\
0.01763	&	0.51682	&	-0.09637	&	-0.45602	&	-0.09657	\\
0.03129	&	0.51698	&	-0.09634	&	-0.45616	&	-0.09655	\\
0.04872	&	0.51774	&	-0.09625	&	-0.45672	&	-0.09645	\\
0.06969	&	0.51965	&	-0.09604	&	-0.45805	&	-0.09622	\\
0.09370	&	0.52344	&	-0.09558	&	-0.46065	&	-0.09569	\\
0.11979	&	0.53016	&	-0.09462	&	-0.46514	&	-0.09467	\\
0.14613	&	0.54120	&	-0.09255	&	-0.47232	&	-0.09260	\\
0.15360	&	0.54569	&	-0.09155	&	-0.47504	&	-0.09181	\\
0.16061	&	0.55081	&	-0.09035	&	-0.47811	&	-0.09077	\\
0.16873	&	0.55852	&	-0.08817	&	-0.48286	&	-0.08916	\\
0.17185	&	0.56268	&	-0.08674	&	-0.48618	&	-0.08855	\\

 \hline

    \end{tabular}
    \caption{Scalar $(l=2)$-led fundamental quasinormal modes with $M_z=2$ for different scaled coupling strength $\xi$ for $j=0.2$.}
    \label{tab:j0.2-l2-sc}
\end{table}

\begin{table}
    \centering
    \begin{tabular}{|c||c|c||c|c|} 
    \hline
  $\alpha/M^2$   & $\omega_R$ & $\omega_I$ & $\omega_R$ & $\omega_I$  \\
 \hline
0.00000	&	0.70858	&	-0.09615	&	-0.64818	&	-0.09626	\\
0.00196	&	0.70855	&	-0.09615	&	-0.64816	&	-0.09626	\\
0.00784	&	0.70854	&	-0.09615	&	-0.64816	&	-0.09626	\\
0.01763	&	0.70880	&	-0.09614	&	-0.64837	&	-0.09625	\\
0.03129	&	0.70982	&	-0.09610	&	-0.64917	&	-0.09618	\\
0.04872	&	0.71227	&	-0.09598	&	-0.65108	&	-0.09601	\\
0.06969	&	0.71707	&	-0.09574	&	-0.65479	&	-0.09567	\\
0.09370	&	0.72529	&	-0.09529	&	-0.66116	&	-0.09504	\\
0.11979	&	0.73819	&	-0.09450	&	-0.67113	&	-0.09396	\\
0.14613	&	0.75738	&	-0.09293	&	-0.68588	&	-0.09207	\\
0.15360	&	0.76465	&	-0.09217	&	-0.69136	&	-0.09123	\\
0.16061	&	0.77275	&	-0.09116	&	-0.69740	&	-0.09019	\\
0.16873	&	0.78489	&	-0.08939	&	-0.70630	&	-0.08827	\\
0.17185	&	0.79122	&	-0.08884	&	-0.71147	&	-0.08740	\\

 \hline

    \end{tabular}
    \caption{Scalar $(l=3)$-led fundamental quasinormal modes with $M_z=2$ for different scaled coupling strength $\xi$ for $j=0.2$.}
    \label{tab:j0.2-l3-sc}
\end{table}


\begin{table}
    \centering
    \begin{tabular}{|c||c|c||c|c||c|c||c|c|}
\hline
  \multirow{2}{*}{}  &
  \multicolumn{2}{|c||}{Axial} &
  \multicolumn{2}{|c||}{Polar} &
  \multicolumn{2}{|c||}{Axial} &
  \multicolumn{2}{|c|}{Polar} \\ 
  $\alpha/M^2$   & $\omega_R$ & $\omega_I$ & $\omega_R$ & $\omega_I$ & $\omega_R$ & $\omega_I$ & $\omega_R$ & $\omega_I$ \\
 \hline
0.00000	&	0.43984	&	-0.08688	&	0.43984	&	-0.08688	&	-0.33246	&	-0.08913	&	-0.33246	&	-0.08913	\\
0.02933	&	0.43997	&	-0.08688	&	0.43942	&	-0.08684	&	-0.33250	&	-0.08915	&	-0.33233	&	-0.08919	\\
0.06548	&	0.44051	&	-0.08692	&	0.43778	&	-0.08673	&	-0.33270	&	-0.08925	&	-0.33177	&	-0.08938	\\
0.10066	&	0.44161	&	-0.08699	&	0.43267	&	-0.08690	&	-0.33315	&	-0.08949	&	-0.33077	&	-0.08983	\\
0.11349	&	0.44220	&	-0.08705	&	0.43089	&	-0.08752	&	-0.33341	&	-0.08964	&	-0.33029	&	-0.09003	\\
0.13978	&	0.44390	&	-0.08737	&	0.42553	&	-0.08887	&	-0.33428	&	-0.09013	&	-0.32909	&	-0.09047	\\
0.15264	&	0.44508	&	-0.08777	&	0.42486	&	-0.09090	&	-0.33508	&	-0.09053	&	-0.32827	&	-0.09063	\\
0.16453	&	0.44654	&	-0.08864	&	0.42636	&	-0.09666	&	-0.33648	&	-0.09097	&	-0.32739	&	-0.09067	\\
0.16871	&	0.44690	&	-0.08928	&	0.42841	&	-0.10406	&	-0.33745	&	-0.09096	&	-0.32696	&	-0.08996	\\
0.17139	&	0.44642	&	-0.09239	&	0.43551	&	-0.11008	&	-0.33472	&	-0.08934	&	-0.32076	&	-0.08949	\\

 \hline

    \end{tabular}
    \caption{ Gravitational $(l=2)$-led fundamental quasinormal modes with $M_z=2$ for different scaled coupling strength $\xi$ for $j=0.4$.}
    \label{tab:j0.4-l2-grav}
\end{table}

\begin{table}
    \centering
    \begin{tabular}{|c||c|c||c|c||c|c||c|c|}
\hline
  \multirow{2}{*}{}  &
  \multicolumn{2}{|c||}{Axial} &
  \multicolumn{2}{|c||}{Polar} &
  \multicolumn{2}{|c||}{Axial} &
  \multicolumn{2}{|c|}{Polar} \\ 
  $\alpha/M^2$   & $\omega_R$ & $\omega_I$ & $\omega_R$ & $\omega_I$ & $\omega_R$ & $\omega_I$ & $\omega_R$ & $\omega_I$ \\
 \hline
0.00000	&	0.66892	&	-0.09062	&	0.66892	&	-0.09062	&	-0.55528	&	-0.09235	&	-0.55528	&	-0.09235	\\
0.02933	&	0.66911	&	-0.09063	&	0.66786	&	-0.09067	&	-0.55537	&	-0.09236	&	-0.55472	&	-0.09244	\\
0.06548	&	0.66998	&	-0.09065	&	0.66347	&	-0.09097	&	-0.55577	&	-0.09240	&	-0.55244	&	-0.09282	\\
0.08830	&	0.67105	&	-0.09068	&	0.65925	&	-0.09126	&	-0.55620	&	-0.09243	&	-0.55017	&	-0.09321	\\
0.10066	&	0.67186	&	-0.09070	&	0.65656	&	-0.09143	&	-0.55657	&	-0.09245	&	-0.54869	&	-0.09345	\\
0.11349	&	0.67295	&	-0.09073	&	0.65348	&	-0.09169	&	-0.55703	&	-0.09249	&	-0.54701	&	-0.09372	\\
0.13978	&	0.67645	&	-0.09079	&	0.64667	&	-0.09244	&	-0.55843	&	-0.09261	&	-0.54306	&	-0.09430	\\
0.15264	&	0.67940	&	-0.09070	&	0.64322	&	-0.09280	&	-0.55952	&	-0.09269	&	-0.54083	&	-0.09453	\\
0.16453	&	0.68362	&	-0.09020	&	0.63992	&	-0.09276	&	-0.56114	&	-0.09276	&	-0.53854	&	-0.09480	\\
0.16871	&	0.68598	&	-0.08994	&	0.63848	&	-0.09241	&	-0.56286	&	-0.09212	&	-0.53751	&	-0.09492	\\
0.17139	&	0.68403	&	-0.08822	&	0.63711	&	-0.09232	&		&		&		&		\\

 \hline

    \end{tabular}
    \caption{ Gravitational $(l=3)$-led fundamental quasinormal modes with $M_z=2$ for different scaled coupling strength $\xi$ for $j=0.4$.}
    \label{tab:j0.4-l3-grav}
\end{table}

\begin{table}
    \centering
    \begin{tabular}{|c||c|c||c|c|} 
    \hline
  $\alpha/M^2$   & $\omega_R$ & $\omega_I$ & $\omega_R$ & $\omega_I$  \\
 \hline
0.00000	&	0.55965	&	-0.09493	&	-0.43306	&	-0.09598	\\
0.02933	&	0.55933	&	-0.09485	&	-0.43299	&	-0.09601	\\
0.06548	&	0.56186	&	-0.09469	&	-0.43432	&	-0.09577	\\
0.08830	&	0.56632	&	-0.09453	&	-0.43632	&	-0.09541	\\
0.10066	&	0.56868	&	-0.09426	&	-0.43757	&	-0.09502	\\
0.11349	&	0.57243	&	-0.09395	&	-0.43927	&	-0.09452	\\
0.13978	&	0.58371	&	-0.09269	&	-0.44425	&	-0.09312	\\
0.15264	&	0.59154	&	-0.09139	&	-0.44776	&	-0.09201	\\
0.16453	&	0.60105	&	-0.08947	&	-0.45201	&	-0.09023	\\
0.16871	&	0.60409	&	-0.08695	&	-0.45377	&	-0.08952	\\
0.17139	&	0.61548	&	-0.08590	&	-0.46267	&	-0.08415	\\

 \hline

    \end{tabular}
    \caption{Scalar $(l=2)$-led fundamental quasinormal modes with $M_z=2$ for different scaled coupling strength $\xi$ for $j=0.4$.}
    \label{tab:j0.4-l2-sc}
\end{table}

\begin{table}
    \centering
    \begin{tabular}{|c||c|c||c|c|} 
    \hline
  $\alpha/M^2$   & $\omega_R$ & $\omega_I$ & $\omega_R$ & $\omega_I$  \\
 \hline
0.00000	&	0.75052	&	-0.09481	&	-0.62547	&	-0.09561	\\
0.02933	&	0.75162	&	-0.09480	&	-0.62620	&	-0.09555	\\
0.06548	&	0.75851	&	-0.09468	&	-0.63039	&	-0.09516	\\
0.08830	&	0.76626	&	-0.09458	&	-0.63515	&	-0.09468	\\
0.10066	&	0.77166	&	-0.09450	&	-0.63851	&	-0.09434	\\
0.11349	&	0.77828	&	-0.09442	&	-0.64264	&	-0.09389	\\
0.13978	&	0.79591	&	-0.09412	&	-0.65377	&	-0.09266	\\
0.15264	&	0.80727	&	-0.09393	&	-0.66100	&	-0.09188	\\
0.16453	&	0.82047	&	-0.09400	&	-0.66970	&	-0.09068	\\
0.16871	&	0.82627	&	-0.09420	&	-0.67351	&	-0.09003	\\
0.17139	&	0.83371	&	-0.09431	&		&		\\

 \hline

    \end{tabular}
    \caption{Scalar $(l=3)$-led fundamental quasinormal modes with $M_z=2$ for different scaled coupling strength $\xi$ for $j=0.4$.}
    \label{tab:j0.4-l3-sc}
\end{table}


\begin{table}
    \centering
    \begin{tabular}{|c||c|c||c|c||c|c||c|c|}
\hline
  \multirow{2}{*}{}  &
  \multicolumn{2}{|c||}{Axial} &
  \multicolumn{2}{|c||}{Polar} &
  \multicolumn{2}{|c||}{Axial} &
  \multicolumn{2}{|c|}{Polar} \\ 
  $\alpha/M^2$   & $\omega_R$ & $\omega_I$ & $\omega_R$ & $\omega_I$ & $\omega_R$ & $\omega_I$ & $\omega_R$ & $\omega_I$ \\
 \hline
0.00000	&	0.49404	&	-0.08377	&	0.49404	&	-0.08377	&	-0.31678	&	-0.08889	&	-0.31678	&	-0.08889	\\
0.01799	&	0.49409	&	-0.08379	&	0.49399	&	-0.08372	&	-0.31673	&	-0.08889	&	-0.31680	&	-0.08890	\\
0.04040	&	0.49429	&	-0.08390	&	0.49309	&	-0.08347	&	-0.31687	&	-0.08893	&	-0.31656	&	-0.08916	\\
0.07139	&	0.49489	&	-0.08428	&	0.49057	&	-0.08316	&	-0.31709	&	-0.08902	&	-0.31610	&	-0.08912	\\
0.08600	&	0.49533	&	-0.08459	&	0.48868	&	-0.08282	&	-0.31726	&	-0.08909	&	-0.31574	&	-0.08930	\\
0.10991	&	0.49632	&	-0.08540	&	0.48442	&	-0.08228	&	-0.31768	&	-0.08925	&	-0.31502	&	-0.08953	\\
0.11832	&	0.49680	&	-0.08583	&	0.48229	&	-0.08222	&	-0.31790	&	-0.08935	&	-0.31464	&	-0.08937	\\
0.12688	&	0.49750	&	-0.08634	&	0.48016	&	-0.08209	&	-0.31820	&	-0.08989	&	-0.31370	&	-0.08824	\\
0.14424	&	0.49630	&	-0.08955	&	0.47480	&	-0.08144	&	-0.31911	&	-0.08911	&	-0.30131	&	-0.08448	\\
0.15285	&	0.49365	&	-0.09319	&	0.48017	&	-0.08099	&	-0.32087	&	-0.08896	&	-0.29871	&	-0.08363	\\

 \hline

    \end{tabular}
    \caption{ Gravitational $(l=2)$-led fundamental quasinormal modes with $M_z=2$ for different scaled coupling strength $\xi$ for $j=0.6$.}
    \label{tab:j0.6-l2-grav}
\end{table}

\begin{table}
    \centering
    \begin{tabular}{|c||c|c||c|c||c|c||c|c|}
\hline
  \multirow{2}{*}{}  &
  \multicolumn{2}{|c||}{Axial} &
  \multicolumn{2}{|c||}{Polar} &
  \multicolumn{2}{|c||}{Axial} &
  \multicolumn{2}{|c|}{Polar} \\ 
  $\alpha/M^2$   & $\omega_R$ & $\omega_I$ & $\omega_R$ & $\omega_I$ & $\omega_R$ & $\omega_I$ & $\omega_R$ & $\omega_I$ \\
 \hline
0.00000	&	0.72280	&	-0.08728	&	0.72280	&	-0.08728	&	-0.53880	&	-0.09164	&	-0.53880	&	-0.09164	\\
0.01799	&	0.72291	&	-0.08730	&	0.72236	&	-0.08725	&	-0.53883	&	-0.09165	&	-0.53862	&	-0.09167	\\
0.04040	&	0.72338	&	-0.08736	&	0.72050	&	-0.08721	&	-0.53895	&	-0.09168	&	-0.53791	&	-0.09177	\\
0.07139	&	0.72477	&	-0.08762	&	0.71559	&	-0.08706	&	-0.53936	&	-0.09169	&	-0.53601	&	-0.09204	\\
0.08600	&	0.72586	&	-0.08782	&	0.71252	&	-0.08692	&	-0.53966	&	-0.09172	&	-0.53477	&	-0.09220	\\
0.10991	&	0.72870	&	-0.08826	&	0.70656	&	-0.08690	&	-0.54034	&	-0.09182	&	-0.53232	&	-0.09252	\\
0.11832	&	0.73014	&	-0.08849	&	0.70426	&	-0.08692	&	-0.54064	&	-0.09191	&	-0.53134	&	-0.09263	\\
0.12688	&	0.73194	&	-0.08874	&	0.70186	&	-0.08693	&	-0.54097	&	-0.09206	&	-0.53026	&	-0.09275	\\
0.13554	&	0.73362	&	-0.08836	&	0.69948	&	-0.08686	&	-0.54131	&	-0.09235	&	-0.52911	&	-0.09285	\\
0.14424	&	0.73913	&	-0.08545	&	0.69746	&	-0.08630	&	-0.54233	&	-0.09310	&	-0.52788	&	-0.09283	\\
0.15285	&	0.75086	&	-0.08267	&	0.69648	&	-0.08257	&	-0.54801	&	-0.09351	&	-0.52690	&	-0.09217	\\

 \hline

    \end{tabular}
    \caption{ Gravitational $(l=3)$-led fundamental quasinormal modes with $M_z=2$ for different scaled coupling strength $\xi$ for $j=0.6$.}
    \label{tab:j0.6-l3-grav}
\end{table}

\begin{table}
    \centering
    \begin{tabular}{|c||c|c||c|c|} 
    \hline
  $\alpha/M^2$   & $\omega_R$ & $\omega_I$ & $\omega_R$ & $\omega_I$  \\
 \hline
0.00000	&	0.61736	&	-0.09125	&	-0.41315	&	-0.09520	\\
0.01799	&	0.61674	&	-0.09115	&	-0.41304	&	-0.09524	\\
0.04040	&	0.61719	&	-0.09107	&	-0.41325	&	-0.09530	\\
0.07139	&	0.62016	&	-0.09114	&	-0.41448	&	-0.09488	\\
0.08600	&	0.62282	&	-0.09113	&	-0.41544	&	-0.09468	\\
0.10991	&	0.62926	&	-0.09112	&	-0.41752	&	-0.09407	\\
0.11832	&	0.63206	&	-0.09125	&	-0.41859	&	-0.09374	\\
0.12688	&	0.63520	&	-0.09154	&	-0.41988	&	-0.09324	\\
0.13554	&	0.63794	&	-0.09274	&	-0.41578	&	-0.09354	\\
0.14424	&	0.63664	&	-0.09862	&		&		\\
0.15285	&	0.64529	&	-0.10742	&	-0.42470	&	-0.09036	\\

 \hline

    \end{tabular}
    \caption{Scalar $(l=2)$-led fundamental quasinormal modes with $M_z=2$ for different scaled coupling strength $\xi$ for $j=0.6$.}
    \label{tab:j0.6-l2-sc}
\end{table}

\begin{table}
    \centering
    \begin{tabular}{|c||c|c||c|c|} 
    \hline
  $\alpha/M^2$   & $\omega_R$ & $\omega_I$ & $\omega_R$ & $\omega_I$  \\
 \hline
0.00000	&	0.80644	&	-0.09150	&	-0.60624	&	-0.09463	\\
0.01799	&	0.80667	&	-0.09151	&	-0.60640	&	-0.09463	\\
0.04040	&	0.80890	&	-0.09163	&	-0.60760	&	-0.09453	\\
0.07139	&	0.81601	&	-0.09199	&	-0.61126	&	-0.09421	\\
0.08600	&	0.82098	&	-0.09220	&	-0.61387	&	-0.09399	\\
0.10991	&	0.83141	&	-0.09273	&	-0.61952	&	-0.09349	\\
0.11832	&	0.83576	&	-0.09299	&	-0.62198	&	-0.09328	\\
0.12688	&	0.84060	&	-0.09335	&	-0.62478	&	-0.09306	\\
0.13554	&	0.84552	&	-0.09428	&	-0.62792	&	-0.09285	\\
0.14424	&	0.84562	&	-0.09915	&		&		\\
0.15285	&	0.81975	&	-0.10347	&	-0.63422	&	-0.09296	\\

 \hline

    \end{tabular}
    \caption{Scalar $(l=3)$-led fundamental quasinormal modes with $M_z=2$ for different scaled coupling strength $\xi$ for $j=0.6$.}
    \label{tab:j0.6-l3-sc}
\end{table}


\begin{table}
    \centering
    \begin{tabular}{|c||c|c||c|c||c|c||c|c|}
\hline
  \multirow{2}{*}{}  &
  \multicolumn{2}{|c||}{Axial} &
  \multicolumn{2}{|c||}{Polar} &
  \multicolumn{2}{|c||}{Axial} &
  \multicolumn{2}{|c|}{Polar} \\ 
  $\alpha/M^2$   & $\omega_R$ & $\omega_I$ & $\omega_R$ & $\omega_I$ & $\omega_R$ & $\omega_I$ & $\omega_R$ & $\omega_I$ \\
 \hline
0.00000	&	0.58602	&	-0.07563	&	0.58602	&	-0.07563	&	-0.30331	&	-0.08851	&	-0.30331	&	-0.08851	\\
0.00800	&	0.58602	&	-0.07563	&	0.58592	&	-0.07560	&	-0.30328	&	-0.08853	&	-0.30330	&	-0.08849	\\
0.01152	&	0.58601	&	-0.07565	&	0.58543	&	-0.07571	&	-0.30324	&	-0.08867	&	-0.30339	&	-0.08852	\\
0.03200	&	0.58605	&	-0.07590	&	0.58528	&	-0.07559	&	-0.30336	&	-0.08853	&	-0.30347	&	-0.08830	\\
0.03872	&	0.58595	&	-0.07593	&	0.58508	&	-0.07552	&	-0.30341	&	-0.08853	&	-0.30360	&	-0.08829	\\
0.04607	&	0.58592	&	-0.07611	&	0.58491	&	-0.07577	&	-0.30332	&	-0.08849	&	-0.30318	&	-0.08785	\\
0.05407	&	0.58578	&	-0.07640	&	0.58351	&	-0.07460	&	-0.30339	&	-0.08866	&	-0.30425	&	-0.08836	\\
0.06270	&	0.58568	&	-0.07671	&	0.58377	&	-0.07240	&	-0.30369	&	-0.08883	&	-0.30717	&	-0.08663	\\
0.06725	&	0.58571	&	-0.07710	&	0.58525	&	-0.07052	&	-0.30403	&	-0.08891	&	-0.30888	&	-0.08319	\\
0.07196	&	0.58578	&	-0.07805	&	0.58796	&	-0.07027	&	-0.30469	&	-0.08888	&	-0.30930	&	-0.07747	\\
0.07683	&	0.58613	&	-0.07996	&	0.59014	&	-0.07137	&	-0.30597	&	-0.08843	&	-0.30709	&	-0.06595	\\
0.08185	&	0.58799	&	-0.08276	&	0.59221	&	-0.07282	&	-0.30828	&	-0.08625	&	-0.31376	&	-0.05827	\\
0.08703	&	0.59228	&	-0.08455	&	0.59566	&	-0.07371	&	-0.30887	&	-0.08088	&	-0.31181	&	-0.05421	\\
0.09236	&	0.59702	&	-0.08448	&	0.60087	&	-0.07058	&	-0.30689	&	-0.07584	&	-0.31010	&	-0.05025	\\
0.09509	&	0.59926	&	-0.08439	&	0.60251	&	-0.06804	&	-0.30539	&	-0.07377	&	-0.31170	&	-0.04963	\\
0.09785	&	0.60129	&	-0.08472	&	0.60384	&	-0.06578	&	-0.30352	&	-0.07201	&	-0.31368	&	-0.05369	\\
0.10348	&	0.60238	&	-0.08585	&	0.60658	&	-0.06229	&	-0.29802	&	-0.06949	&	-0.31163	&	-0.05601	\\
0.10926	&	0.60176	&	-0.08430	&	0.60993	&	-0.06127	&	-0.28709	&	-0.06751	&	-0.30675	&	-0.04668	\\
0.11221	&	0.60160	&	-0.08314	&	0.61105	&	-0.06271	&	-0.27783	&	-0.06444	&	-0.30145	&	-0.04049	\\

 \hline

    \end{tabular}
    \caption{ Gravitational $(l=2)$-led fundamental quasinormal modes with $M_z=2$ for different scaled coupling strength $\xi$ for $j=0.8$.}
    \label{tab:j0.8-l2-grav}
\end{table}

\begin{table}
    \centering
    \begin{tabular}{|c||c|c||c|c||c|c||c|c|}
\hline
  \multirow{2}{*}{}  &
  \multicolumn{2}{|c||}{Axial} &
  \multicolumn{2}{|c||}{Polar} &
  \multicolumn{2}{|c||}{Axial} &
  \multicolumn{2}{|c|}{Polar} \\ 
  $\alpha/M^2$   & $\omega_R$ & $\omega_I$ & $\omega_R$ & $\omega_I$ & $\omega_R$ & $\omega_I$ & $\omega_R$ & $\omega_I$ \\
 \hline
0.00000	&	0.80678	&	-0.07905	&	0.80678	&	-0.07905	&	-0.52505	&	-0.09061	&	-0.52505	&	-0.09061	\\
0.00800	&	0.80682	&	-0.07906	&	0.80658	&	-0.07942	&	-0.52504	&	-0.09059	&	-0.52502	&	-0.09062	\\
0.01152	&	0.80685	&	-0.07906	&	0.80658	&	-0.07900	&	-0.52505	&	-0.09055	&	-0.52504	&	-0.09060	\\
0.03200	&	0.80736	&	-0.07923	&	0.80526	&	-0.07887	&	-0.52507	&	-0.09054	&	-0.52460	&	-0.09065	\\
0.03872	&	0.80766	&	-0.07930	&	0.80464	&	-0.07880	&	-0.52511	&	-0.09036	&	-0.52443	&	-0.09066	\\
0.04607	&	0.80811	&	-0.07943	&	0.80384	&	-0.07878	&	-0.52596	&	-0.09051	&	-0.52421	&	-0.09068	\\
0.05407	&	0.80865	&	-0.07982	&	0.80268	&	-0.07902	&	-0.52667	&	-0.09097	&	-0.52404	&	-0.09096	\\
0.06270	&	0.80910	&	-0.08084	&	0.80084	&	-0.07893	&	-0.52712	&	-0.09213	&	-0.52312	&	-0.09103	\\
0.06725	&	0.80892	&	-0.08169	&	0.80003	&	-0.07886	&	-0.52791	&	-0.09332	&	-0.52281	&	-0.09095	\\
0.07196	&	0.80777	&	-0.08269	&	0.79936	&	-0.07899	&	-0.52894	&	-0.09493	&	-0.52251	&	-0.09074	\\
0.07683	&	0.80406	&	-0.08349	&	0.79874	&	-0.08017	&	-0.53055	&	-0.09710	&	-0.52238	&	-0.09052	\\
0.08185	&	0.79761	&	-0.09028	&	0.79562	&	-0.08189	&	-0.53306	&	-0.09940	&	-0.52238	&	-0.09034	\\
0.08703	&	0.79921	&	-0.09005	&	0.79087	&	-0.08096	&	-0.53624	&	-0.10072	&	-0.52245	&	-0.09018	\\
0.09236	&	0.79955	&	-0.08743	&	0.79018	&	-0.07624	&	-0.53833	&	-0.10127	&	-0.52262	&	-0.09006	\\
0.09509	&	0.80027	&	-0.08580	&	0.79062	&	-0.07585	&	-0.53915	&	-0.10241	&	-0.52282	&	-0.09004	\\
0.09785	&	0.80165	&	-0.08402	&	0.79070	&	-0.07561	&	-0.54151	&	-0.10353	&	-0.52315	&	-0.09008	\\
0.10348	&	0.80691	&	-0.08007	&	0.79060	&	-0.07521	&	-0.54284	&	-0.10168	&	-0.52433	&	-0.09032	\\
0.10926	&	0.81497	&	-0.07484	&	0.79022	&	-0.07521	&	-0.54147	&	-0.10213	&	-0.52455	&	-0.09034	\\
0.11221	&	0.81858	&	-0.07159	&	0.78942	&	-0.07549	&	-0.54146	&	-0.10087	&	-0.52407	&	-0.09013	\\

 \hline

    \end{tabular}
    \caption{ Gravitational $(l=3)$-led fundamental quasinormal modes with $M_z=2$ for different scaled coupling strength $\xi$ for $j=0.8$.}
    \label{tab:j0.8-l3-grav}
\end{table}

\begin{table}
    \centering
    \begin{tabular}{|c||c|c||c|c|} 
    \hline
  $\alpha/M^2$   & $\omega_R$ & $\omega_I$ & $\omega_R$ & $\omega_I$  \\
 \hline
0.00000	&	0.70682	&	-0.08152	&	-0.39573	&	-0.09429	\\
0.00800	&	0.70628	&	-0.08146	&	-0.39568	&	-0.09431	\\
0.01152	&	0.70600	&	-0.08156	&	-0.39566	&	-0.09430	\\
0.03872	&	0.70536	&	-0.08169	&	-0.39579	&	-0.09432	\\
0.04607	&	0.70570	&	-0.08154	&	-0.39571	&	-0.09445	\\
0.05407	&	0.70550	&	-0.08113	&	-0.39602	&	-0.09361	\\
0.06270	&	0.70548	&	-0.08141	&	-0.39584	&	-0.09376	\\
0.06725	&	0.70551	&	-0.08165	&	-0.39553	&	-0.09314	\\
0.07196	&	0.70558	&	-0.08197	&	-0.39569	&	-0.09174	\\
0.07683	&	0.70605	&	-0.08258	&	-0.39726	&	-0.09013	\\
0.08185	&	0.70885	&	-0.08435	&	-0.39999	&	-0.08993	\\
0.08703	&	0.71532	&	-0.08279	&	-0.40207	&	-0.09150	\\
0.09236	&	0.71865	&	-0.08071	&	-0.40262	&	-0.09371	\\
0.09509	&	0.72027	&	-0.08054	&	-0.40262	&	-0.09481	\\
0.09785	&	0.72189	&	-0.08036	&	-0.40293	&	-0.09581	\\
0.10348	&	0.72462	&	-0.07817	&	-0.40502	&	-0.09641	\\
0.10926	&	0.72825	&	-0.07111	&	-0.40404	&	-0.09695	\\
0.11221	&	0.73147	&	-0.06804	&	-0.40642	&	-0.09250	\\

 \hline

    \end{tabular}
    \caption{Scalar $(l=2)$-led fundamental quasinormal modes with $M_z=2$ for different scaled coupling strength $\xi$ for $j=0.8$.}
    \label{tab:j0.8-l2-sc}
\end{table}

\begin{table}
    \centering
    \begin{tabular}{|c||c|c||c|c|} 
    \hline
  $\alpha/M^2$   & $\omega_R$ & $\omega_I$ & $\omega_R$ & $\omega_I$  \\
 \hline
0.00000	&	0.88948	&	-0.08309	&	-0.58983	&	-0.09336	\\
0.00800	&	0.88943	&	-0.08308	&	-0.58982	&	-0.09336	\\
0.01152	&	0.88947	&	-0.08308	&	-0.58984	&	-0.09336	\\
0.03200	&	0.89060	&	-0.08341	&	-0.59046	&	-0.09332	\\
0.03872	&	0.89098	&	-0.08300	&	-0.59083	&	-0.09330	\\
0.04607	&	0.89193	&	-0.08323	&	-0.59134	&	-0.09330	\\
0.05407	&	0.89311	&	-0.08329	&	-0.59198	&	-0.09325	\\
0.06270	&	0.89455	&	-0.08325	&	-0.59283	&	-0.09320	\\
0.06725	&	0.89541	&	-0.08312	&	-0.59334	&	-0.09317	\\
0.07196	&	0.89640	&	-0.08288	&	-0.59391	&	-0.09314	\\
0.07683	&	0.89755	&	-0.08249	&	-0.59453	&	-0.09313	\\
0.08185	&	0.89891	&	-0.08189	&	-0.59520	&	-0.09315	\\
0.08703	&	0.90056	&	-0.08099	&	-0.59582	&	-0.09323	\\
0.09236	&	0.90259	&	-0.07957	&	-0.59609	&	-0.09332	\\
0.09509	&	0.90381	&	-0.07854	&	-0.59585	&	-0.09321	\\
0.09785	&	0.90523	&	-0.07717	&	-0.59523	&	-0.09266	\\
0.10348	&	0.90888	&	-0.07274	&	-0.59368	&	-0.08899	\\
0.10926	&	0.91344	&	-0.06379	&	-0.59151	&	-0.08761	\\
0.11221	&	0.91475	&	-0.05660	&	-0.59199	&	-0.09133	\\

 \hline

    \end{tabular}
    \caption{Scalar $(l=3)$-led fundamental quasinormal modes with $M_z=2$ for different scaled coupling strength $\xi$ for $j=0.8$.}
    \label{tab:j0.8-l3-sc}
\end{table}

\end{document}